\documentclass[aps,prd]{revtex4}

\usepackage{graphicx}
\usepackage{amsmath}
\usepackage{array}
\usepackage{feynmf}

\def\lg{{\mathchoice{~\raise.58ex\hbox{$<$}\mkern-14.8mu\lower.52ex\hbox{$>$}~}
                    {~\raise.58ex\hbox{$<$}\mkern-14.8mu\lower.52ex\hbox{$>$}~}
                    {\raise.59ex\hbox{{$\scriptscriptstyle <$}}\mkern-12.8mu%
                     \lower.01ex\hbox{{$\scriptscriptstyle >$}}}   {}   }}
\def\gl{{\mathchoice{~\raise.58ex\hbox{$>$}\mkern-12.8mu\lower.52ex\hbox{$<$}~}
                    {~\raise.58ex\hbox{$>$}\mkern-12.8mu\lower.52ex\hbox{$<$}~}
                    {\raise.62ex\hbox{{$\scriptscriptstyle >$}}\mkern-12.0mu%
                     \lower.05ex\hbox{{$\scriptscriptstyle <$}}}  {}    }}

\newcommand{\be}{\begin{equation}}
\newcommand{\ee}{\end{equation}}
\newcommand{\ba}{\begin{eqnarray}}
\newcommand{\ea}{\end{eqnarray}}
\newcommand{\ban}{\begin{eqnarray*}}
\newcommand{\ean}{\end{eqnarray*}}
\newcommand \nn {\nonumber}
\newcommand{\sla}{\!\!\!/ \,}

\begin{document}

\title{${\cal N} =4$ Super Yang-Mills Plasma}

\author{Alina Czajka}

\affiliation{Institute of Physics, Jan Kochanowski University, Kielce, Poland}

\author{Stanis\l aw Mr\' owczy\' nski}

\affiliation{Institute of Physics, Jan Kochanowski University, Kielce, Poland}
\affiliation{National Centre for Nuclear Research, Warsaw, Poland}

\date{November 10, 2014}

\begin{abstract}

The ${\cal N} =4$ super Yang-Mills plasma is studied in the regime of weak coupling. Collective excitations and collisional processes are discussed. Since the Keldysh-Schwinger approach is used, the collective excitations in both equilibrium and non-equilibrium plasma are under consideration. The dispersion equations of gluon, fermion, and scalar fields are written down and the self-energies, which enter the equations, are computed in the Hard Loop Approximation. The self-energies are discussed in the context of effective action which is also given. The gluon modes and fermion ones appear to be the same as those in the QCD plasma of gluons and massless quarks. The scalar modes are as of a free relativistic massive particle. The binary collisional processes, which occur at the lowest nontrivial order of the coupling constant, are reviewed and then the transport properties of the plasma are discussed. The ${\cal N} =4$ super Yang-Mills plasma is finally concluded to be very similar to the QCD plasma of gluons and light quarks. The differences mostly reflect different numbers of degrees of freedom in the two systems. 

\end{abstract}



\maketitle

\section{Introduction}

Supersymmetric models are considered as possible extensions of the Standard Model, see {\it e.g.} \cite{Signer:2009dx}, and supersymmetry is then assumed to be a symmetry of Nature at a sufficiently high energy scale. Experiments at the Large Hadron Collider may soon show whether this is the case.  This paper is devoted to a plasma system with the dynamics governed by the ${\cal N} =4$ supersymmetric Yang-Mills theory \cite{Brink:1976bc,Gliozzi:1976qd}.  In the models with an extended (${\cal N} > 1$) supersymmetry, the left- and right-handed fermions interact in the same way, in conflict with the Standard Model where the left- and right-handed matter particles are coupled differently.  Consequently, the ${\cal N} =4$ super Yang-Mills is not treated as a serious candidate for a theory to describe the world of elementary particles. Nevertheless, the theory attracts a lot of attention because of its unique features. The ${\cal N} =4$ super Yang-Mills appears to be finite and thus it is conformally invariant not only at the classical but at the quantum level as well. 

A great interest in the ${\cal N} =4$ super Yang-Mills theory was stimulated by a discovery of the AdS/CFT duality of the five-dimensional gravity in the anti de Sitter geometry and the conformal field theories \cite{Maldacena:1997re}, for a review see \cite{Aharony:1999ti} and the lecture notes \cite{Klebanov:2000me} as an introduction. The duality offered a unique tool to study strongly coupled field theories. Since the gravitational constant and the coupling constant of dual conformal field theory are inversely proportional to each other, some problems of strongly coupled field theories can be solved via weakly coupled gravity. In this way some intriguing features of strongly coupled systems driven by the ${\cal N} =4$ super Yang-Mills dynamics were revealed, see the reviews \cite{Son:2007vk,Janik:2010we}, but the relevance of the results for non-supersymmetric systems, which are of our actual interest, remains an open issue. In particular, one asks how properties of the ${\cal N} =4$ super Yang-Mills plasma (SYMP) are related to those of quark-gluon plasma (QGP) studied experimentally in relativistic heavy-ion collisions. While such a comparison is, in general, a difficult problem, some comparative analyses have been done in the domain of weak coupling where perturbative methods are applicable \cite{CaronHuot:2006te,Huot:2006ys,CaronHuot:2008uh,Blaizot:2006tk,Chesler:2006gr,Chesler:2009yg}.

We undertook a task of systematic comparison of supersymmetric perturbative plasmas to their non-supersymmetric counterparts. We started with the ${\cal N} =1$ SUSY QED, analyzing first collective excitations of ultrarelativistic plasma which, in general, is out of equilibrium \cite{Czajka:2010zh}. We computed the one-loop retarded self-energies of photons, photinos, electrons and selectrons in the Hard Loop Approximation using the Keldysh-Schwinger formalism. The self-energies, which we also analyzed in the context of effective action, enter the dispersion equations of photons, photinos, electrons and selectrons, respectively. The collective modes of ${\cal N} =1$ SUSY QED plasma appear to be essentially the same as  those in ultrarelativistic electromagnetic plasma of photons, electrons and positrons. In particular, a spectrum of photino modes coincides with that of quasi-electrons. Therefore, independent of whether photon modes are stable or unstable, there are no unstable photino excitations. The supersymmetry, which is obviously broken in the plasma medium, does not induce any instability in the photino sector.

In the subsequent paper \cite{Czajka:2011zn} we discussed collisional characteristics of ${\cal N} =1$ SUSY QED plasma. For this purpose we computed cross sections of all elementary processes which occur at the lowest non-trivial order of $\alpha\equiv e^2/4\pi$. We found that some processes, {\it e.g.} the Compton scattering on selectrons, are independent of momentum transfer. The processes are qualitatively different from those of the usual electromagnetic interactions dominated by small momentum transfers.  Further on we discussed collisional characteristics of equilibrium ${\cal N} =1$ SUSY QED plasma, observing that parameters of ultrarelativistic plasmas are strongly constrained by dimensional arguments, as the temperature is the only dimensional quantity of equilibrium system. Then, transport coefficients like viscosity are proportional to appropriate powers of temperature and the coefficients characterizing different plasmas can differ only by numerical factors. So, we derived  the energy loss and momentum broadening of a particle traversing the equilibrium plasma, which depend not only on the plasma temperature but on the energy of the test particle as well. We found that the two quantities have very similar structure (in the limit of the high energy of the test particle) even for very different elementary cross sections. Our findings presented in  \cite{Czajka:2010zh,Czajka:2011zn} show that the plasmas of ${\cal N} =1$ SUSY QED and of QED are surprisingly similar to each other.  In this paper we discuss properties of the ${\cal N} =4$ super Yang-Mills plasma, analyzing both collective excitations and collisional characteristics of the system. 

Our main aim is to confront the weakly coupled plasma driven by ${\cal N} =4$ super Yang-Mills with  the perturbative quark-gluon plasma governed by QCD. We do not attempt to compare our results to those obtained in a strong coupling regime using either the AdS/CFT duality or lattice QCD. Some plasma characteristics we discuss, {\it e.g.} the energy loss, are computed in both strongly and weakly coupled systems but it is rather unclear how to study collective excitations representing colored quasiparticles in the setting of AdS/CFT duality or lattice QCD. The paper \cite{Bak:2007fk} demonstrates that even the definition of Debye screening mass, which has a very  simple meaning in perturbative plasmas,  is not straightforward in strongly interacting systems. For these reasons we escape from discussing our results in the context of strong coupling.

Our paper is organized as follows. In the next section,  we discuss the Lagrangian of ${\cal N} = 4$ super Yang-Mills and the field content of the system under consideration. The vertices of ${\cal N} = 4$ super Yang-Mills are collected in Appendix \ref{sec-SYM-vertex}.  In Sec.~\ref{sec-basic} basic characteristics of SYMP such as energy density and Debye mass are discussed and compared to those of QGP. Then, we move to plasma collective excitations. The general dispersion equations of gauge bosons, fermions and scalars are written down in Sec.~\ref{sec-dis-eqs} and the self-energies, which enter the equations, are obtained in the subsequent section. We apply here the Keldysh-Schwinger approach which allows one to study equilibrium and non-equilibrium systems. The free Green's functions of Keldysh-Schwinger formalism are given in Appendix \ref{sec-Green-fun}. Since we are interested in collective modes, the self-energies are obtained in the long wavelength limit corresponding to the Hard Loop Approximation. The effective action of the Hard Loop Approach is derived in Sec.~\ref{sec-eff-action} and possible structures of self-energies are considered in this context. In Sec.~\ref{sec-modes} we present a qualitative discussion of collective modes in SYMP. Sec.~\ref{sec-collisions} is devoted to collisional characteristics of the plasma - elementary processes and transport coefficients are briefly discussed here. Finally, we conclude our study in Sec.~\ref{sec-conclusions}. 

As we have intended to make our paper complete and self-contained, there is inevitably some repetition of the content of our previous publications \cite{Czajka:2010zh,Czajka:2011zn}, mostly in Secs.~\ref{sec-modes}, \ref{sec-collisions}. Throughout the paper we use the natural system of units with $c= \hbar = k_B =1$; our choice of the signature of the metric tensor is $(+ - - -)$.

\section{${\cal N}=4$ Super Yang-Mills Theory}
\label{sec-lagrangian}

We start our considerations with a discussion of the Lagrangian of ${\cal N}=4$ super Yang-Mills theory \cite{Brink:1976bc,Gliozzi:1976qd}. We follow here the presentation given in \cite{Yamada:2006rx}.

The gauge group is assumed to be ${\rm SU}(N_c)$ and every field of the ${\cal N}=4$ super Yang-Mills theory belongs to its  adjoint representation.  The field content of the theory, which is summarized in Table~\ref{table-field-content}, is the following. There are gauge bosons (gluons) described by the vector field $A_\mu^a$ with $a, b, c, \dots = 1, 2, \dots N_c^2 -1$. There are four Majorana fermions represented by the Weyl spinors $\lambda^\alpha$ with $\alpha = 1,2$ which can be combined in the Dirac bispinors as
\be
\label{Majorana-bispinor}
\Psi  = \left(
   \begin{matrix}
   \lambda^\alpha  \cr
   \bar{\lambda}_{\dot{\alpha}}  \cr
   \end{matrix}
   \right) ,
\;\;\;\;\;\;\;
\bar{\Psi}  = (  \lambda_\alpha,  \bar{\lambda}^{\dot{\alpha}}  ) ,
\ee
where $\bar{\lambda}_{\dot{\alpha}} \equiv [ \lambda_\alpha]^\dagger$ with $\dagger$ denoting Hermitian conjugation. To numerate the Majorana fermions we use the indices $i, j = 1,2,3,4$ and the corresponding bispinor is denoted as $\Psi_i$. Finally, there are six real scalar fields which are assembled in the multiplet $\Phi = (X_1, Y_1, X_2, Y_2, X_3, Y_3)$. The components of $\Phi$ are either denoted as $X_p$ for scalars, and $Y_p$ for pseudoscalars, with $p,q =1,2,3$ or as $\Phi_A$ with $A, B=1,2, \dots 6$.

\begin{table}[b]
\caption{\label{table-field-content} Field content of ${\cal N} =4$ super Yang-Mills theory.}
\begin{ruledtabular}
\begin{tabular}{ccccc}
Field's symbol &  Type of the field & Range of the field's index & Spin & Number
of degrees of freedom $(N^{\rm dof})$
\\[4mm]
$A^\mu$ & vector &$\mu, \nu = 0,1,2,3$ & 1& $2 \times (N_c^2-1)$
\\[2mm]
$\Phi_A$ & real (pseudo-)scalar & $A, B = 1,2, 3,4,5, 6$ & 0 & $6 \times (N_c^2-1)$
\\[2mm]
$\lambda_i$ & Majorana spinor & $i, j = 1,2,3,4$ & 1/2 & $8 \times (N_c^2-1)$
\end{tabular}
\end{ruledtabular}
\end{table}

The Lagrangian density of ${\cal N} =4$ super Yang-Mills theory can be written as
\ba
\label{Lagrangian-1}
{\cal L}
&=&
-\frac{1}{4}F^{\mu \nu}_a F_{\mu \nu}^a
+\frac{i}{2}\bar \Psi_i^a (D\!\sla \Psi_i)^a
+\frac{1}{2}(D_\mu \Phi_A)_a (D^\mu \Phi_A)_a
\\ [2mm] \nn
&&
-\frac{1}{4} g^2f^{abe} f^{cde} \Phi_A^a \Phi_B^b \Phi_A^c \Phi_B^d
-i\frac{g}{2} f^{abc} \Big( \bar \Psi_i^a  \alpha_{ij}^p  X_p^b \Psi_j^c  
+i\bar \Psi_i^a \beta_{ij}^p\gamma_5  Y_p^b \Psi_j^c \Big),
\ea
where
$F^{\mu \nu}_a = \partial^\mu A^\nu_a - \partial^\nu A^\mu_a + g f^{abc} A^\mu_b A^\nu_c$ and the covariant derivatives equal $(D\!\sla \Psi_i)^a = (\partial \,\!\sla \delta_{ab} +g f^{abc} A_c \!\! \sla) \Psi_i^b$ and $(D^\mu \Phi)_a = D^\mu_{ab} \Phi_b = (\partial^\mu \delta_{ab} + gf^{abc}A^\mu_c)\Phi_b$; $g$ is the coupling constant; $f^{abc}$ are the structure constants of the ${\rm SU}(N_c)$ group; and the $4 \times 4$ matrices $\alpha^p, \beta^p$ satisfy the relations
\be
\label{alpha-beta-relations}
\{\alpha^p, \alpha^q \} = - 2 \delta^{p q},
\;\;\;\;\;\;\;
\{\beta^p, \beta^q \} = - 2 \delta^{p q},
\;\;\;\;\;\;\;
[ \alpha^p, \beta^q] = 0 ,
\ee
and their explicit form can be chosen as
\ba
\label{alphas}
\alpha^1  &=& \left(
\begin{matrix}
0 & \sigma_1  \cr
- \sigma_1 & 0  \cr
\end{matrix}
   \right) ,
\;\;\;\;\;\;\;
\alpha^2  = \left(
\begin{matrix}
0 & -\sigma_3  \cr
\sigma_3 & 0  \cr
\end{matrix}
   \right) ,
\;\;\;\;\;\;\;
\alpha^3  = \left(
\begin{matrix}
i \sigma_2 & 0  \cr
0 & i \sigma_2  \cr
\end{matrix}
   \right) ,
\\[2mm]
\label{betas}
\beta^1  &=& \left(
\begin{matrix}
0 & i\sigma_2  \cr
i \sigma_2 & 0  \cr
\end{matrix}
   \right) ,
\;\;\;\;\;\;\;
\beta^2  = \left(
\begin{matrix}
0 & \sigma_0  \cr
-\sigma_0 & 0  \cr
\end{matrix}
   \right) ,
\;\;\;\;\;\;\;
\beta^3  = \left(
\begin{matrix}
-i \sigma_2 & 0  \cr
0 & i \sigma_2  \cr
\end{matrix}
   \right) ,
\ea
where the $2 \times 2$ Pauli matrices read
\ba
\label{Pauli}
\sigma^0  = \left(
\begin{matrix}
1 & 0  \cr
0 & 1  \cr
\end{matrix}
   \right) ,
\;\;\;\;\;\;\;
\sigma^1  = \left(
\begin{matrix}
0 & 1  \cr
1 & 0  \cr
\end{matrix}
   \right) ,
\;\;\;\;\;\;\;
\sigma^2  = \left(
\begin{matrix}
0 & -i  \cr
i & 0  \cr
\end{matrix}
   \right) ,
\;\;\;\;\;\;\;
\sigma^3  = \left(
\begin{matrix}
1 & 0  \cr
0 & -1  \cr
\end{matrix}
   \right) .
\ea
As seen, the matrices $\alpha^p, \beta^p$  are antiHermitian: $(\alpha^p)^\dagger = -\alpha^p$ , $(\beta^p)^\dagger = -\beta^p$. The vertices of ${\cal N} = 4$ super Yang-Mills, which can be inferred from the Lagrangian (\ref{Lagrangian-1}),  are collected in Appendix \ref{sec-SYM-vertex}. The vertices are used in perturbative calculations presented in the subsequent sections.

The Lagrangian (\ref{Lagrangian-1}) is sometimes \cite{CaronHuot:2008uh,Chesler:2006gr,Chesler:2009yg} extended by adding a fundamental ${\cal N} = 2$  hypermultiplet to mimic a behavior of quarks in QCD plasma. The hypermultiplet is typically massive to study heavy flavors but it can be massless as well. We do not consider any extension of the Lagrangian (\ref{Lagrangian-1}) but at the end of Sec.~\ref{sec-eff-action} we briefly comment on a possible structure of self-energies of fields belonging to the fundamental ${\cal N} = 2$  hypermultiplet.

\section{Basic plasma characteristics}
\label{sec-basic}

We start our discussion of SYMP with basic characteristics of the equilibrium plasma. Specifically, we consider the energy and particle densities, Debye mass and plasma parameter of SYMP comparing the quantities to those of QGP. For the beginning, however, a few comments are in order. 

In QGP there are several conserved charges: baryon number, electric and color charges, strangeness. The net baryon number and electric charge are typically non-zero in  QGP produced in relativistic heavy-ion collisions while the total strangeness and color charge vanish. Actually, the color charge is usually assumed to vanish not only globally but locally as well. It certainly makes sense as the whitening of QGP appears to be the relaxation process of the shortest time scale \cite{Manuel:2004gk}. In SYMP, there are  conserved charges carried by fermions and scalars associated with the global ${\rm SU}(4)$ symmetry. One of these charges can be identified with the electric charge to couple ${\cal N} = 4$ super Yang-Mills to the electromagnetic field \cite{CaronHuot:2006te}. In the forthcoming the average ${\rm SU}(4)$ charges of SYMP are assumed to vanish and so are the associated chemical potentials. The constituents of SYMP carry color charges but we further assume that the plasma is globally and locally colorless. 

Since there are conserved supercharges in supersymmetric theories, it seems reasonable to consider a statistical supersymmetric system with a non-zero expectation value of the supercharge. However, it is not obvious how to deal with a partition function customary defined as ${\rm Tr}e^{-\beta (H - \mu Q)}$ where $\beta \equiv T^{-1}$ is the inverse temperature, $H$ is the Hamiltonian, $Q$ is the supercharge operator and $\mu$ is the associated chemical potential. The problem is caused by a fermionic character of the supercharge $Q$. If $\mu$ is simply a number, as, say, the baryon chemical  potential, the partition function even of non-interacting system does not factorize into a product of partition functions of single momentum modes because the supercharges of different modes do not commute with each other. The supercharge is not an extensive quantity \cite{Kapusta:1984cp}. There were proposed two ways to resolve the problem. Either the chemical  potential remains a number but the supercharge is modified by multiplying it by an additional fermionic field $c$ \cite{Kapusta:1984cp,Mrowczynski:1986cu} or the chemical  potential by itself is a fermionic field \cite{Kovtun:2003vj}. Then, $\mu c Q$ and $\mu Q$ are both bosonic and the partition function can be computed in a standard way. The two formulations, however, are not equivalent to each other. According to the former one \cite{Kapusta:1984cp,Mrowczynski:1986cu}, properties of a supercharged system vary with an expectation value of the supercharge, within the latter one  \cite{Kovtun:2003vj}, the partition function appears to be effectively independent of $Q$. Because of the ambiguity, we further consider SYMP where the expectation values of all supercharges vanish  both globally and locally.

In view of the above discussion, SYMP is comparable to QGP where the conserved charges are all zero and so are the associated chemical potentials. We adopt the assumption whenever the two plasma systems are compared to each other. 

When the chemical potentials are absent, the temperature $(T)$ is the only dimensional parameter, which characterizes the equilibrium plasma, and all plasma parameters are expressed through the appropriate powers of $T$. Taking into account the right numbers of bosonic and fermionic degrees of freedom in SYMP and QGP, the energy densities of equilibrium non-interacting plasmas equal
\be
\varepsilon = \frac{\pi^2}{60} \, {30 (N_c^2 -1)  \choose 4(N_c^2 -1) + 7 N_f N_c}  T^4.
\ee
where the upper expression is for SYMP and the lower one for QGP with $N_f$ light quark flavors. The quark is {\em light} when its mass is much smaller than the plasma temperature. For $N_c=N_f =3$, the energy density of SYMP is approximately 2.5 times bigger than that of QGP at the same temperature. The same holds for the pressure $p$ which, obviously, equals $\varepsilon/3$.

The particle densities in SYMP and QGP are found to be
\be
n = \frac{2\zeta(3)}{\pi^2} \, {7 (N_c^2 -1)  \choose 2(N_c^2 -1) + 3 N_f N_c}  T^3,
\ee
where $\zeta(3) \approx 1.202$ is the Riemann zeta function. For $N_c=N_f =3$ we have $n_{\rm SYMP}/n_{\rm QGP} \approx 1.3$ at the same temperature.

As we show in Sec.~\ref{subsec-polar-tensor}, the gluon polarization tensor has exactly the same structure in SYMP and QGP, and consequently the Debye mass in SYMP is defined in the same way as in QGP. The masses in both plasmas equal 
\be
m_D^2 = \frac{g^2}{6} \, {12 N_c \choose  2 N_c  + N_f}   T^2,
\ee
where, as previously, the upper case is for SYMP and the lower one for QGP. For $N_c=N_f =3$, the ratio of Debye masses squared is 2.4 at the same value of $gT$. The Debye mass determines not only the screening length $r_D = 1/m_D$ but it also gives the plasma frequency  $\omega_p = m_D/\sqrt{3}$ which is the minimal frequency of longitudinal and transverse plasma oscillations corresponding to the zero wave vector. The  plasma frequency  is also called the gluon thermal mass. 

Another important quantity characterizing the equilibrium plasma is the so-called plasma parameter $\lambda$ which equals the inverse number of particles in the sphere of radius of the screening length. When $\lambda$ is decreasing, the behavior of plasma is more and more collective while  inter-particle collisions are less and less important. For $N_c=N_f =3$, we have
\be
\lambda \equiv \frac{1}{\frac{4}{3} \pi r_D^3 n} \approx {0.257 \choose 0.042} g^3 .
\ee
As seen, the dynamics of QGP is more collective than that of SYMP.

The differences of $\varepsilon$ and $n$ for SYMP and QGP merely reflect the difference in numbers of degrees of freedom in the two plasma systems. In the case of $m_D$ and $\lambda$ it also matters that (anti-)quarks in QGP and fermions in SYMP  belong to different representations - fundamental and adjoint, respectively - of the ${\rm SU}(N_c)$ gauge group. 

\section{Dispersion equations}
\label{sec-dis-eqs}

Dispersion equations determine dispersion relations of quasi-particle excitations. Below
we write down the dispersion equations of quasi-gluons, quasi-fermions, and quasi-scalars.

\subsection{Gluons}

Since the equation of motion of the gluon field $A^{\mu}_a(k)$  can be written in the form
\be
\label{eq-motion-A}
\Big[ k^2 g^{\mu \nu} -k^{\mu} k^{\nu} - \Pi^{\mu \nu}(k) \Big]
A_{\nu}(k) = 0 ,
\ee
where color indices are dropped, $\Pi^{\mu \nu}(k)$ is the retarded polarization tensor and $k\equiv (\omega, {\bf k})$ is the four-momentum, the general gluon dispersion equation is
\be
\label{dis-photon-1}
{\rm det}\Big[ k^2 g^{\mu \nu} -k^{\mu} k^{\nu} - \Pi^{\mu \nu}(k) \Big]
 = 0 .
\ee
Strictly speaking, one should consider the equation of motion not of the gluon field but of the gluon propagator. Then, Eq.~(\ref{dis-photon-1}) determines the poles of the propagator. Because of the transversality of $\Pi^{\mu \nu}$ ($k_\mu \Pi^{\mu \nu}(k) =0$), which is required by the gauge covariance, not all components of $\Pi^{\mu \nu}$ are independent from each other, and consequently the dispersion equation (\ref{dis-photon-1})  can be much simplified by expressing the polarization tensor through the dielectric tensor $\varepsilon^{ij}(k)$ which is the $3 \times 3$ not $4 \times 4$ matrix.

\subsection{Fermions}

The fermion field $\psi_i (k)$ obeys the equation
\be
\Big[ k\sla  - \Sigma (k)  \Big] \psi (k) =0 ,
\ee
where any indices are neglected and $\Sigma (k)$ is the retarded fermion self-energy.
The dispersion equation thus is
\be
\label{dis-electron-1}
 {\rm det}\Big[ k\sla  - \Sigma (k) \Big]  = 0 .
\ee
Further on we assume that the spinor structure of
$\Sigma(k)$ is
\be
\label{structure-electron}
\Sigma (k) = \gamma^{\mu} \Sigma_{\mu}(k) .
\ee
Then, substituting the expression (\ref{structure-electron}) into Eq.~(\ref{dis-electron-1}) and computing the determinant as explained in Appendix 1 of \cite{Mrowczynski:1992hq}, we get
\be
\label{dis-electron-2}
\Big[\big( k^{\mu} - \Sigma^{\mu}(k)
\big) \big(k_{\mu} - \Sigma_{\mu}(k) \big)\Big]^2  = 0 .
\ee

\subsection{Scalars}

The scalar field $\Phi_A (k)$ obeys the Klein-Gordon equation
\be
\label{dis-eq-selectron}
\Big[ k^2  + P(k) \Big] \Phi(k) =0 ,
\ee
where $P(k)$ is the retarded self-energy of the scalar field and any indices are dropped. The dispersion equation is
\be
\label{dis-selectron}
k^2 + P(k) = 0 .
 \ee

As seen, the whole dynamical information about plasma medium is contained in the self-energies which are computed perturbatively in the next section. 

\section{Self-energies}
\label{sec-self-energies}

We compute here the self-energies which enter the dispersion equations (\ref{dis-photon-1}, \ref{dis-electron-1}, \ref{dis-selectron}).  The vertices of ${\cal N} = 4$ super Yang-Mills, which are used in our perturbative calculations,  are listed in Appendix \ref{sec-SYM-vertex}. The plasma is assumed to be homogeneous (translationally invariant), locally colorless but the momentum distribution is, in general, different from the equilibrium one. Therefore, we adopt the Keldysh-Schwinger or real-time formalism which allows one to describe both equilibrium and non-equilibrium many-body systems. The free Green's functions, which are labeled with the indices $+,-, >, <, {\rm sym}$, are collected in Appendix \ref{sec-Green-fun}. What concerns the Keldysh-Schwinger  formalism we follow the conventions explained in \cite{Mrowczynski:1992hq}. The computation is performed within the Hard Loop Approach, see the reviews  \cite{Thoma:1995ju,Blaizot:2001nr,Kraemmer:2003gd}, which was generalized to anisotropic systems in \cite{Mrowczynski:2000ed}. 

\subsection{Polarization tensor}
\label{subsec-polar-tensor}

\begin{figure}[t]
\centering
\includegraphics*[width=0.7\textwidth]{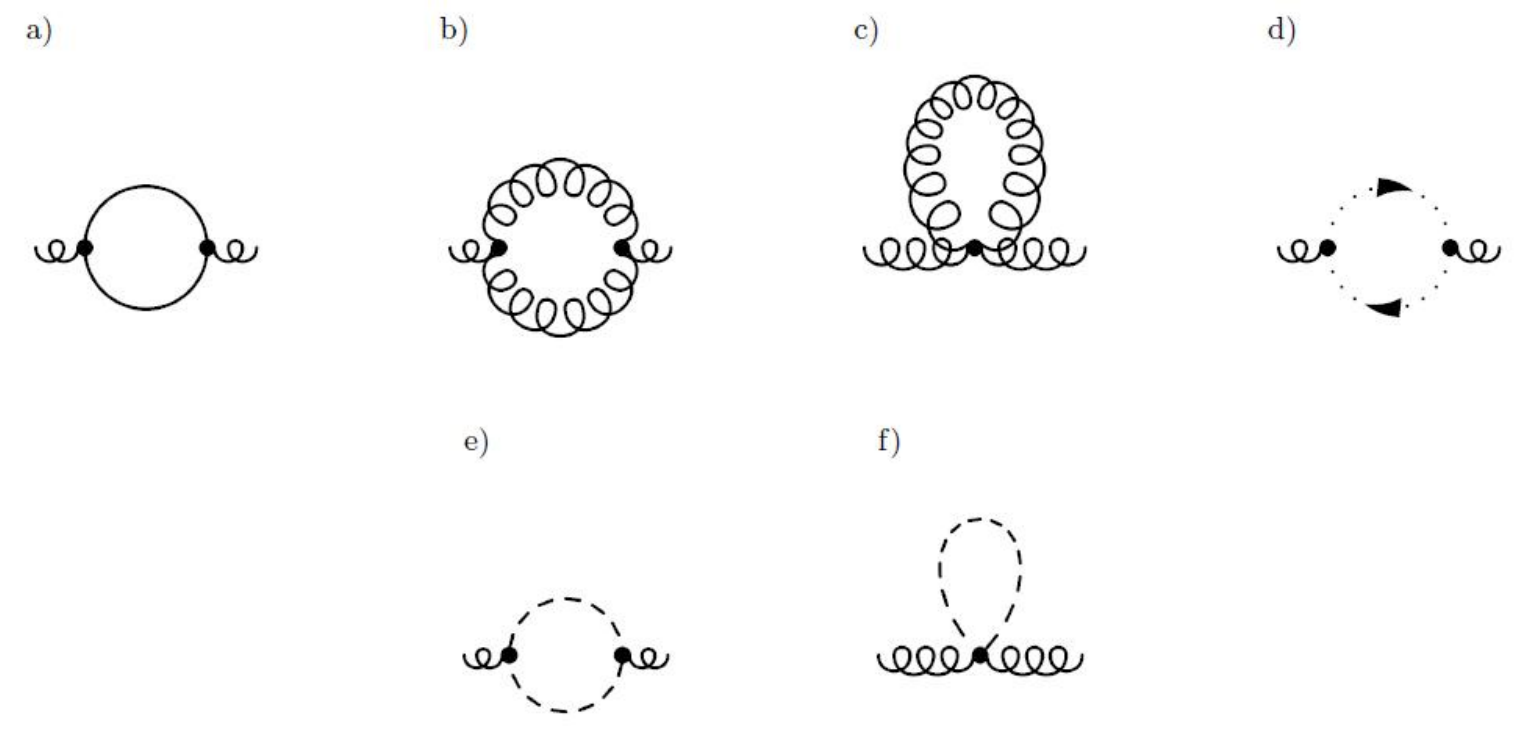}
\caption{Contributions to the gluon self-energy. }
\label{fig-gluon}
\end{figure}

The gluon polarization tensor $\Pi^{\mu \nu}$ can be defined by means of the Dyson-Schwinger equation
\be
i{\cal D}^{\mu \nu} (k) = i D^{\mu \nu} (k)
+ i D^{\mu \rho}(k) \, i\Pi_{\rho \sigma}(k)  \, i{\cal D}^{\sigma \nu}(k) ,
\ee
where ${\cal D}^{\mu \nu}$ and $D^{\mu \nu}$ are the interacting and free gluon propagator, respectively. The lowest order contributions to gluon polarization tensor are given by six diagrams shown in Fig.~\ref{fig-gluon}. The curly, plain, dotted and dashed lines denote, respectively, gluon, fermion, ghost, and scalar fields.

Using the vertices given in Appendix \ref{sec-SYM-vertex}, the contribution to the {\em contour} polarization tensor of Keldysh-Schwinger formalism, which comes from the fermion loop corresponding to the graph in Fig.~\ref{fig-gluon}a, is immediately written down in the coordinate space as
\be
\label{contour-Pi}
_{(a)}\Pi^{\mu \nu}_{ab}(x,y)
= -ig^2 N_c \delta_{ab}
{\rm Tr} [\gamma^\mu S_{ij}(x,y) \gamma^\nu S_{ji}(y,x)] .
\ee
where the trace is taken over spinor indices. The factor $(-1)$ due to the fermion loop is included and the relation $f_{acd }f_{bcd}=\delta_{ab} N_c$ is used here. 

We are interested in the retarded polarization tensor which is expressed
through $\Pi^\lg$ as
\be
\Pi^+ (x,y) =  \Theta(x_0 - y_0)
\Big( \Pi^> (x,y) - \Pi^< (x,y) \Big).
\ee
The polarization tensors $\Pi^\lg$ are found from the contour tensor (\ref{contour-Pi}) by locating the argument $x_0$ on the upper (lower) and $y_0$ on the lower (upper) branch of the contour. Then, one gets
\be
\label{Pi->-<-x-y}
\big( {_{(a)} \Pi^{\lg} (x,y)} \big)^{\mu \nu}_{ab}
= -ig^2 N_c \delta_{ab}
{\rm Tr} [\gamma^\mu S^\lg_{ij} (x,y) \gamma^\nu S^\gl_{ji} (y,x)] .
\ee

As already mentioned, the system under study is assumed to be translationally invariant. Then, the two-point functions as $S(x,y)$ effectively depend on $x$ and $y$ only through $x-y$. Therefore, we put $y=0$ and we write $S(x,y)$ as $S(x)$ and $S(y,x)$ as $S(-x)$. Then, Eq.~(\ref{Pi->-<-x-y}) is
\be
\label{Pi->-<-x-y-1}
\big(  {_{(a)} \Pi^{\lg} (x)} \big)^{\mu \nu}_{ab}
= -\frac{i}{2} g^2 N_c \delta_{ab}
{\rm Tr} [\gamma^\mu S^\lg_{ij} (x) \gamma^\nu S^\gl_{ji} (-x)] .
\ee
Since
\be
S^{\pm} (x) = \pm \Theta(\pm x_0)
\Big(S^> (x) - S^< (x) \Big)
\ee
the retarded polarization tensor $\Pi^+ (x)$ is found as
\be
\label{Pi-x-1}
\big({_{(a)} \Pi^+(x)} \big)^{\mu \nu}_{ab}
 = -\frac{i}{2} g^2 N_c \delta_{ab}
{\rm Tr}\big[\gamma^\mu S^+_{ij}(x) \gamma^\nu S^{\rm sym}_{ji}(-x)
+ \gamma^\mu S^{\rm sym}_{ji}(x) \gamma^\nu S^-_{ij}(-x) \big],
\ee
which in the momentum space reads
\be
\label{Pi-k-e-1}
\big({_{(a)} \Pi^+(k)} \big)^{\mu \nu}_{ab}
= -\frac{i}{2} g^2 N_c \delta_{ab}
 \int \frac{d^4p}{(2\pi)^4}
{\rm Tr} \big[\gamma^\mu S^+_{ij}(p+k) \gamma ^\nu S^{\rm sym}_{ji}(p)
+ \gamma^\mu S^{\rm sym}_{ji}(p) \gamma^\nu S^-_{ij}(p-k) \big].
\ee

Further on the index $+$ is dropped and $\Pi^+$ is denoted as  $\Pi$, as only the {\em retarded} polarization tensor is discussed. Substituting the functions $S^{\pm}, S^{\rm sym}$ given by Eqs.~(\ref{S-pm}, \ref{S-<}, \ref{S->}) into the formula (\ref{Pi-k-e-1}), one finds
\ba
\label{Pi-k-e-4}
 _{(a)}\Pi^{\mu \nu}_{ab}(k)
&=& - 4 g^2 N_c  \delta_{ab}
\int \frac{d^3p}{(2\pi)^3} \, \frac{2n_f({\bf p})-1}{E_p}
\\ \nonumber
&& \times
\bigg(
\frac{2p^\mu p^\nu + k^\mu p^\nu + p^\mu k^\nu - g^{\mu \nu} (k \cdot p)}
{(p+k)^2 + i\, {\rm sgn}\big((p+k)_0\big)0^+}
+ \frac{2p^\mu p^\nu - k^\mu p^\nu - p^\mu k^\nu + g^{\mu \nu} (k \cdot p)}
{(p-k)^2 - i\, {\rm sgn}\big((p-k)_0\big)0^+}
\bigg) ,
\ea
where $p^\mu \equiv (E_p, {\bf p})$ with $E_p \equiv |{\bf p}|$,  the traces of gamma matrices are computed and it is taken into account that $p^2 =0$. We also note that after performing the integration over $p_0$, the momentum ${\bf p}$ was changed into  $-{\bf p}$ in the negative energy contribution. 

In the Hard Loop Approximation, when $p \gg k$, we have
\ba
\label{HLA-plus}
\frac{1}{(p+k)^2 + i0^+}
+ \frac{1} {(p-k)^2 - i0^+}
&=& \frac{2k^2}{(k^2)^2 - 4 (k\cdot p)^2 - i {\rm sgn}(k\cdot p) 0^+}
\approx -\frac{1}{2}  \frac{k^2}{(k\cdot p + i 0^+)^2} ,
\\ [2mm]
\label{HLA-minus}
\frac{1}{(p+k)^2 + i0^+}
- \frac{1} {(p-k)^2 - i0^+}
&=& \frac{4(k \cdot p)}{(k^2)^2 - 4 (k\cdot p)^2 - i {\rm sgn}(k\cdot p) 0^+}
\approx \frac{k\cdot p}{(k\cdot p + i 0^+)^2}.
\ea
We note that $(p+k)_0 > 0$ and $(p-k)_0 > 0$ for $p \gg k$. With the formulas (\ref{HLA-plus}, \ref{HLA-minus}), Eq.~(\ref{Pi-k-e-4}) gives
\ba
\label{Pi-k-e-final}
 _{(a)} \Pi^{\mu \nu}_{ab}(k)
 &=& 4 g^2 N_c \delta_{ab}
 \int \frac{d^3p}{(2\pi)^3} \, \frac{2n_f({\bf p})-1}{E_p} \,
\frac{k^2 p^\mu p^\nu - \big(k^\mu p^\nu + p^\mu k^\nu
- g^{\mu \nu} (k \cdot p) \big) (k \cdot p)}
{(k\cdot p + i 0^+)^2} ,
\ea
which has the well-known structure of the polarization tensor of gauge bosons in ultrarelativistic QED and QCD plasmas.  As seen, $\Pi(k)$ is symmetric with respect to Lorentz indices $ {_{(a)}\Pi}^{\mu \nu}_{ab}(k) =  {_{(a)}\Pi}^{\nu \mu}_{ab}(k)$ and transverse $k_\mu  {_{(a)} \Pi}^{\mu \nu}_{ab}(k) = 0$, as required by the gauge invariance. In the vacuum limit, when the fermion distribution function $n_f({\bf p})$ vanishes, the polarization tensor (\ref{Pi-k-e-final}) is still nonzero (actually infinite). As we will see, the vacuum contribution to the complete polarization tensor exactly vanishes due to the supersymmetry.

In analogy to the fermion-loop expression (\ref{Pi-k-e-1}), one finds the gluon-loop contribution to the retarded polarization tensor shown in Fig.~\ref{fig-gluon}b as
\ba
\nonumber
 _{(b)}\Pi^{\mu \nu}_{ab}(k) &=& - i\frac{g^2}{4} N_c \delta_{ab}
 \int \frac{d^4p}{(2\pi )^4}  \int \frac{d^4q}{(2\pi )^4} D^{\rm sym}(p)
\Big[ (2\pi)^4 \delta^{(4)}(k+p-q)
M^{\mu \nu} (k,q,p) D^+(q)
\\ [2mm]
\label{Pi-gluon-loop-2}
&& \;\;\;\;\;\;\;\;\;\;\;\;\;\;\;\;\;\;\;\;\;\;\;\;\;\;\;\;\;\;\;\;\;\;\;\;\;\;\;\;\;\;\;\;\;\;\;\;\;\;\;\;\;\;\;\;\;\;
+ (2\pi)^4 \delta^{(4)}(k-p+q)
M^{\mu \nu} (k,-q,-p) D^-(q)
\Big],
\ea
where the gluon Green's functions $D^\pm$ and $D^{\rm sym}$ are given by Eqs.~(\ref{D-pm}, \ref{D-sym}), the combinatorial factor $1/2$ is included and 
\be
\label{tensor-M-def}
M^{\mu \nu} (k,q,p) \equiv
\Gamma^{\mu \sigma \rho} (k,-q,p)
\Gamma^{\;\;\,\nu}_{\sigma \;\; \rho} (q,-k,-p) 
\ee
with 
\be
\label{3-g-vertex-2}
\Gamma^{\mu \nu \rho} (k,p,q) \equiv
g^{\mu \nu }(k-p)^\rho
+g^{\nu \rho}(p-q)^\mu +g^{\rho \mu}(q-k)^\nu .
\ee

Within the Hard Loop Approximation the tensor (\ref{tensor-M-def}) is computed as
\be
\label{tensor-M-F-HL}
M^{\mu \nu} (k,p \pm k, \pm p) \approx \pm 2 g^{\mu \nu} (k\cdot p)
+ 10 p^\mu p^\nu
\pm 5(k^\mu p^\nu + p^\mu k^\nu),
\ee
where we have taken into account that $p^2=0$.

Substituting the expressions (\ref{tensor-M-F-HL}) into Eq.~(\ref{Pi-gluon-loop-2}), using  the explicit form of the functions $D^\pm$ and $D^{\rm sym}$ given by Eqs.~(\ref{D-pm}, \ref{D-sym}), and applying the Hard Loop Approximation (\ref{HLA-plus}, \ref{HLA-minus}) we get
\ba
\label{Pi-gluon-loop-5}
 _{(b)}\Pi^{\mu \nu}_{ab}(k) =
\frac{g^2}{4} N_c \delta_{ab}
 \int \frac{d^3p}{(2\pi )^3} \frac{2n_g({\bf p})+1}{E_p}
\frac{5k^2 p^\mu p^\nu - 2 g^{\mu \nu} (k\cdot p)^2
- 5(k^\mu p^\nu + p^\mu k^\nu)(k\cdot p)}{(k\cdot p + i 0^+)^2} .
\ea

The gluon-tadpole contribution to the retarded polarization tensor, which shown in Fig.~\ref{fig-gluon}c,
equals
\be
\label{Pi-gluon-tadpole-1}
_{(c)}\Pi^{\mu \nu}_{ab}(k) = - \frac{g^2}{2}
 \int \frac{d^4p}{(2\pi )^4}
\Gamma^{\mu \nu \rho}_{abcc \rho} D^<(p)  ,
\ee
where the combinatorial factor $1/2$ is included and $\Gamma^{\mu \nu \rho \sigma }_{abcd}$ equals
\be
\label{4-g-vertex}
\Gamma^{\mu \nu \rho \sigma }_{abcd} \equiv
f_{abe}f_{ecd}(g^{\mu \sigma} g^{\nu \rho} - g^{\mu \rho} g^{\nu \sigma})
+ f_{ace}f_{edb}(g^{\mu \rho} g^{\nu \sigma} - g^{\mu \nu} g^{\rho \sigma})
+ f_{ade}f_{ebc}(g^{\mu \nu} g^{\rho \sigma} - g^{\mu \sigma} g^{\nu \rho}).
\ee
With the explicit form of the function $D^<(p)$ given by Eq.~(\ref{D-<}), the formula (\ref{Pi-gluon-tadpole-1}) provides 
\be
_{(c)}\Pi^{\mu \nu}_{ab}(k) = \frac{3}{2} g^2 N_c \, \delta_{ab} g^{\mu \nu}
\int \frac{d^3p}{(2\pi )^3} \frac{2 n_g({\bf p}) +1}{E_p} .
\ee

The ghost-loop contribution to the retarded polarization tensor, which is shown in Fig.~\ref{fig-gluon}d, equals
\ba
\label{Pi-ghost-loop-2}
 _{(d)}\Pi^{\mu \nu}_{ab}(k) &=&  i\frac{g^2}{2} N_c \delta_{ab}
 \int \frac{d^4p}{(2\pi )^4} \;G^{\rm sym}(p)
\Big[ (p+k)^\mu p^{\nu} G^+(p+k)
+ p^\mu (p-k)^\nu G^-(p-k) \Big].
\ea
where the factor $(-1)$  is included as we deal with the fermion loop. Using the explicit form of the functions $G^\pm$ and $G^{\rm sym}$ given by Eqs.~(\ref{G-pm}, \ref{G-sym}), the formula (\ref{Pi-ghost-loop-2}) is manipulated to
\be
\label{Pi-ghost-loop-4}
 _{(d)}\Pi^{\mu \nu}_{ab}(k) =  -\frac{g^2}{4} N_c \delta_{ab}
\int \frac{d^3p}{(2\pi )^3} \; \frac{2n_g({\bf p})+1}{E_p}
\frac{k^2 p^\mu p^\nu - (k^\mu p^\nu + p^\mu k^\nu) (k\cdot p)}{(k\cdot p + i 0^+)^2}  .
\ee
which holds in the Hard Loop Approximation.

As already mentioned, the quark-loop contribution to the polarization tensor is symmetric and transverse with respect to Lorentz indices. The same holds for the sum of gluon-loop, gluon-tadpole and ghost-loop contributions which gives the gluon polarization tensor in pure gluodynamics (QCD with no quarks).  The sum of the three contributions equals
\be
\label{Pi-b-c-d}
 _{(b)+(c)+(d)}\Pi^{\mu \nu}_{ab}(k)
= g^2 N_c \delta_{ab}
 \int \frac{d^3p}{(2\pi )^3} \frac{2n_g({\bf p})+ 1}{E_p}
\frac{k^2 p^\mu p^\nu + g^{\mu \nu} (k\cdot p)^2
- (k^\mu p^\nu + p^\mu k^\nu) (k\cdot p)}{(k\cdot p + i 0^+)^2}.
\ee
To our best knowledge this is the first computation of the QCD polarization tensor in Hard Loop Approximation performed in the Keldysh-Schwinger (real time) formalism which explicitly demonstrates the transversality of the tensor. In Refs.~\cite{Weldon:1982aq,Mrowczynski:2000ed}, where the equilibrium and non-equilibrium anisotropic plasmas were considered, respectively, the transversality of $\Pi^{\mu \nu}(k)$ was actually assumed.  In the case of  imaginary time formalism, the computation of  the gluon polarization tensor in Hard Loop Approximation  is the textbook material \cite{lebellac,Kapusta-Gale}. 

The contribution to the polarization tensor coming from the scalar loop depicted
in Fig.~\ref{fig-gluon}e is given by
\be
\label{Pi-s-l-1}
 _{(e)}\Pi^{\mu \nu}_{ab}(k) = - i \frac{g^2}{2} \delta_{ab} N_c \delta^{AA}
\int \frac{d^4p}{(2\pi)^4}
\big[(2p+k)^\mu (2p+k)^\nu  \Delta^+ (p+k)  \Delta^{\rm sym}(p)
+ (2p-k)^\mu (2p-k)^\nu  \Delta^{\rm sym} (p) \Delta^-(p-k) \big],
\ee
which changes into 
\be
\label{Pi-s-l-3}
_{(e)}\Pi^{\mu \nu}_{ab}(k) =  3g^2 N_c \delta_{ab}
\int \frac{d^3p}{(2\pi )^3} \,
\frac{2n_s({\bf p})+1}{E_p} \,
\frac{k^2 p^\mu p^\nu - (p^\mu k^\nu + k^\mu p^\nu)(k \cdot p)}{(k \cdot p + i0^+)^2}.
\ee
when the functions $\Delta^{\pm}$ and $\Delta^{\rm sym}$ given by Eqs.~(\ref{Del-pm}, \ref{Del-sym}) are used and the Hard Loop Approximation is adopted.

The contribution to the polarization tensor coming from the scalar tadpole depicted
in Fig.~\ref{fig-gluon}f is
\be
\label{Pi-s-t-1}
_{(f)}\Pi^{\mu \nu}_{ab}(k) = - \frac{1}{2} 2ig^2 \delta^{ab} N_c \delta^{AA} g^{\mu \nu}
\int \frac{d^4p}{(2\pi)^4}  \Delta^<(p) ,
\ee
where the combinatorial factor $1/2$ is included. With the function $\Delta^<$ given by
Eq.~(\ref{Del-<}) we have
\be
\label{Pi-s-t-2}
_{(f)}\Pi^{\mu \nu}_{ab}(k) = 3 g^2 N_c \delta_{ab} g^{\mu \nu}
\int \frac{d^3p}{(2\pi )^3}  \,
\frac{2n_s({\bf p})+1}{E_p}  .
\ee

We get the complete contribution from a scalar field to the polarization tensor by summing up the scalar loop and scalar tadpole. Thus, one finds
\be
\label{Pi-s-total}
_{(e+f)}\Pi^{\mu \nu}_{ab}(k) =  3 g^2 N_c \delta_{ab}
\int \frac{d^3p}{(2\pi )^3} \,
\frac{2n_s({\bf p})+1}{E_p} \,
\frac{k^2 p^\mu p^\nu - \big(p^\mu k^\nu + k^\mu p^\nu - g^{\mu \nu} (k \cdot p)\big)(k \cdot p)}
{(k \cdot p + i0^+)^2},
\ee
which has the structure corresponding to the scalar QED.  Then, it is not a surprise that the polarization tensor (\ref{Pi-s-total}) is symmetric and transverse. 

After summing up all contributions, we get the final expression of gluon polarization tensor
\be
\label{Pi-k-final}
\Pi^{\mu \nu}_{ab}(k)
= g^2 N_c \delta_{ab}
\int \frac{d^3p}{(2\pi)^3}
\frac{f({\bf p})}{E_p} 
\frac{k^2 p^\mu p^\nu - (k^\mu p^\nu + p^\mu k^\nu - g^{\mu \nu} (k\cdot p))
(k\cdot p)}{(k\cdot p + i 0^+)^2},
\ee
where 
\be
\label{f-def}
f({\bf p}) \equiv 2n_g({\bf p}) + 8n_f({\bf p}) + 6n_s({\bf p})
\ee
is the effective distribution function of plasma constituents.  We observe that the coefficients in front of the distributions functions $n_g({\bf p})$,   $n_f({\bf p})$, $n_s({\bf p})$ equal the numbers of degrees of freedom (except colors) of, respectively, gauge bosons, fermions and scalars, {\it cf.} Table \ref{table-field-content}. This is obviously a manifestation of supersymmetry. Another effect of the supersymmetry is vanishing of the tensor (\ref{Pi-k-final}) in the vacuum limit when $f({\bf p})  = 0$. Needless to say, the polarization tensor (\ref{Pi-k-final}) is symmetric and transverse in Lorentz indices and thus it is gauge independent.

In the case of QCD plasma, one gets the polarization tensor of the form (\ref{Pi-k-final}) after the vacuum contribution is subtracted. For the QGP with the number $N_f$ of massless flavors, the effective distribution function equals
\be
\label{f-def-QGP}
f_{\rm QGP}({\bf p}) \equiv 2n_g({\bf p}) +  \frac{N_f}{N_c} \big(n_q({\bf p}) + n_{\bar q}({\bf p}) \big) ,
\ee
where $n_q({\bf p})$, $n_{\bar q}({\bf p})$ are the distribution functions of quarks and antiquarks which contribute differently to the polarization tensor than fermions of the ${\cal N} = 4$ super Yang-Mills. This happens because (anti-)quarks of QCD belong to the fundamental representation of ${\rm SU}(N_c)$ while the fermions belong to the adjoint representation. 

\subsection{Fermion self-energy}

\begin{figure}[t]
\centering
\includegraphics*[width=0.7\textwidth]{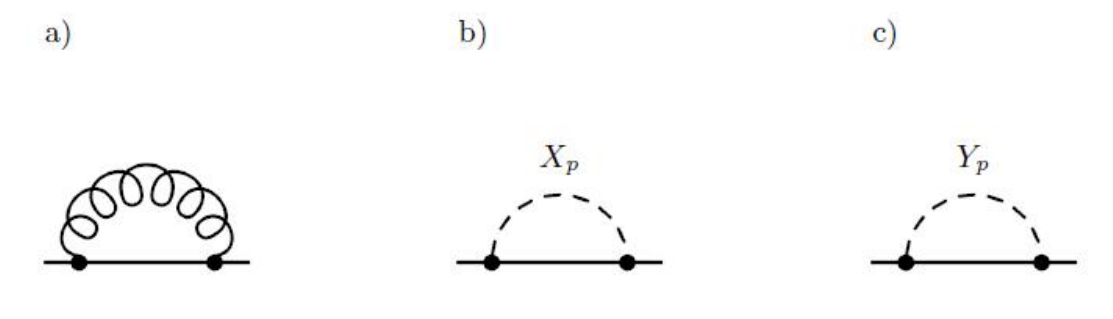}
\caption{Contributions to the fermion self-energy. }
\label{fig-fermion}
\end{figure}

The fermion self-energy $\Sigma$ can be defined by means of the Dyson-Schwinger
equation
\be
i{\cal S} (k) = i S (k) + i S(k) \, \big(-i\Sigma (k) \big) \, i{\cal S}(k) ,
\ee
where ${\cal S}$ and $S$ are the interacting and free propagator, respectively. The lowest order contributions to fermion self-energy are given by diagrams shown in Fig.~\ref{fig-fermion}. The curly, plain, and dashed lines denote, respectively, gluon, fermion, and scalar fields.

The contribution to the fermion self-energy corresponding to the graph depicted in Fig.~\ref{fig-fermion}a is given by
\be
\label{Si-k-a-1}
_{(a)}\Sigma^{ij}_{ab}(k) = \frac{i}{2} g^2
N_c \delta_{ab} \delta^{ij}
\int \frac{d^4p}{(2\pi )^4}
\big[\gamma_\mu S^+(p+k) \gamma^\mu D^{\rm sym}(p)
+ \gamma_\mu S^{\rm sym}(p) \gamma^\mu D^-(p-k)
\big].
\ee
With the functions $D^{\pm}$, $D^{\rm sym}$ and $S^{\pm}$, $S^{\rm sym}$ given by Eqs.~(\ref{D-pm}, \ref{D-sym}, \ref{S-pm}, \ref{S-sym}), one obtains
\ba
\label{Si-k-a-3}
_{(a)}\Sigma^{ij}_{ab}(k) &=& g^2 N_c \delta_{ab} \delta^{ij}
\int \frac{d^3p}{(2\pi )^3} \, \frac{ n_g ({\bf p}) +  n_f ({\bf p})}{E_p}  \,
\frac{p\sla}{k\cdot p + i 0^+} ,
\ea
where the traces over gamma matrices are computed and the Hard Loop Approximation is applied.
Eq.~(\ref{Si-k-a-3}) has the well-known form of electron self-energy in QED.

Since there are scalar and pseudoscalar  fields $X_p$ and $Y_p$, there are two contributions to the fermion self-energy corresponding to the graphs depicted in Figs.~\ref{fig-fermion}b, \ref{fig-fermion}c. The first one corresponding to the $X_p$ field equals
\be
\label{Si-k-b-1}
{_{(b)}\Sigma}^{ij}_{ab}(k) =  i \frac{g^2}{2} N_c \delta^{ab} \alpha^p_{ik}\alpha^p_{kj}
\int \frac{d^4p}{(2\pi )^4}
\big[ S^+(p+k)  \Delta^{\rm sym}(p) + S^{\rm sym}(p)  \Delta^-(p-k) \big].
\ee
Because of the relations (\ref{alpha-beta-relations}), one finds that $\alpha^p_{ik}\alpha^p_{kj}=-3\delta_{ij}$.
Using the result and substituting the functions $S^{\pm}$, $S^{\rm sym}$ and $\Delta^{\pm}$, $\Delta^{\rm sym}$ given by Eqs.~(\ref{S-pm}, \ref{S-sym}, \ref{Del-pm}, \ref{Del-sym}) into Eq.~(\ref{Si-k-b-1}), one obtains the following result
\ba
\label{Si-k-b-2}
{_{(b)}\Sigma}^{ij}_{ab} (k) &=& \frac{3}{2} g^2 N_c \delta_{ab} \delta^{ij}
\int \frac{d^3p}{(2\pi)^3}
\frac{ n_f ({\bf p}) +  n_s ({\bf p})}{E_p}  \,
\frac{p\sla}{k\cdot p + i 0^+} ,
\ea
which holds  in the Hard Loop Approximation.

The contribution due to the pseudoscalar field $Y_p$ is
\be
\label{Si-k-c-1}
{_{(c)}\Sigma}^{ij}_{ab}(k) =  i\frac{g^2}{2} N_c \delta^{ab} \beta^p_{ik}\beta^p_{kj}
\int \frac{d^4p}{(2\pi )^4} \big[ \gamma_5 S^+(p+k) \gamma_5 \Delta^{\rm sym}(p) 
+ \gamma_5 S^{\rm sym}(p) \gamma_5 \Delta^-(p-k) \big].
\ee
Because $\beta^p_{ik}\beta^p_{kj}=-3\delta_{ij}$, $\gamma_\mu \gamma_5=-\gamma_5 \gamma_\mu$, and $\gamma_5^2 = 1$, we again obtain the result (\ref{Si-k-b-2}).

Summing up all the contributions, we get the final expression for the fermion self-energy
\ba
\label{Si-k-final}
\Sigma^{ij}_{ab}(k) &=& \frac{g^2}{2} \, N_c \delta_{ab}\delta^{ij}
\int \frac{d^3p}{(2\pi )^3}
\frac{f({\bf p})}{E_p}  \, \frac{p\sla}{k\cdot p + i 0^+}.
\ea
which, as the polarization tensor (\ref{Pi-k-final}), depends on the effective distribution function (\ref{f-def}).

\subsection{Scalar self-energy}

\begin{figure}[t]
\centering
\includegraphics*[width=0.7\textwidth]{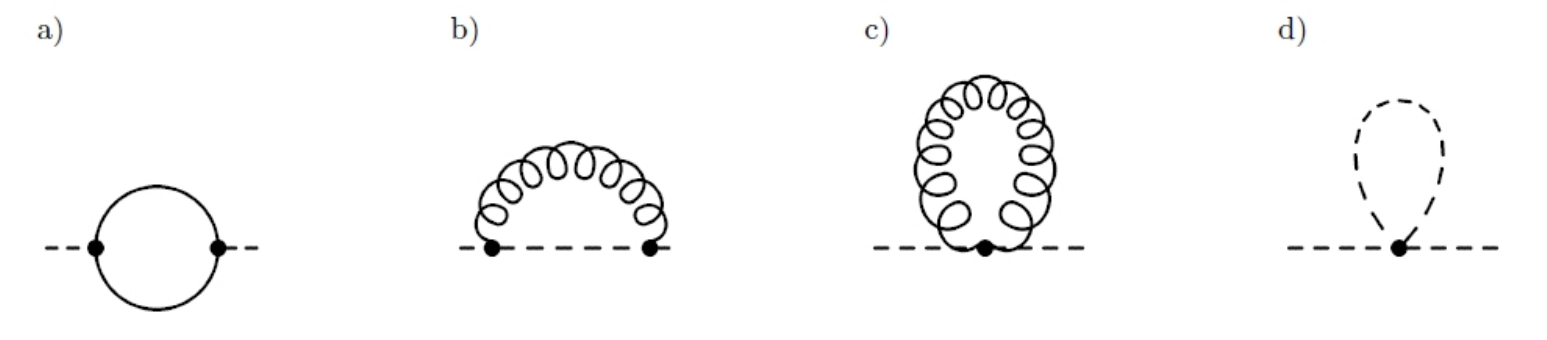}
\caption{Contributions to the scalar self-energy. }
\label{fig-scalar}
\end{figure}

The scalar self-energy $P(k)$ can be defined by means of the Dyson-Schwinger equation
\be
i \tilde \Delta (k) = i \Delta(k)
+ i \Delta (k) \, i P(k)  \, i \tilde\Delta(k) ,
\ee
where $\tilde\Delta$ and $\Delta$ are the scalar interacting and free propagator, respectively. The lowest order contributions to the scalar self-energy are given by the diagrams shown in Fig.~\ref{fig-scalar}. The curly, plain, and dashed lines denote, respectively, gluon, fermion, and scalar fields.

Since there are scalar ($X_p$) and pseudoscalar ($Y_p$) fields, we have to consider separately the self-energies of  $X_p$ and $Y_p$. However, one observes that only the coupling of scalars to fermions differs for $X_p$ and $Y_p$. The self-interaction and the coupling to the gauge field are the same. Therefore, only the fermion-loop contribution to the scalar self-energy, which is shown in the diagram Fig.~\ref{fig-scalar}a, needs to be computed separately for the $X_p$ and $Y_p$ fields.

In the case of the scalar $X_p$ field, the diagram  Fig.~\ref{fig-scalar}a provides
\be
\label{P-k-a-1}
_{(a)} P^{pq}_{ab}(k) =  i \frac{g^2}{4} \, N_c \delta^{ab} \alpha^p_{ij} \alpha^q_{ji} 
\int \frac{d^4p}{(2\pi)^4}
{\rm Tr}\big[   S^+ (p+k)  S^{\rm sym} (p)
+ S^{\rm sym} (p)   S^- (p-k) \big].
\ee
where the symmetry factor $1/2$ and the extra minus sign due to the fermionic character of the loop are included. With the explicit form of the functions $S^{\pm}, \; S^{\rm sym}$  given by Eqs.~(\ref{S-pm}, \ref{S-sym}) and the identity
$\alpha^p_{ij}\alpha^q_{ji}=-4\delta^{pq}$ which follows from the relations (\ref{alpha-beta-relations}), one finds
\be
\label{P-k-a-2}
_{(a)}P^{pq}_{ab}(k) = -4 g^2 N_c \delta_{ab} \delta^{pq}
\int \frac{d^3p}{(2\pi )^3} \, \frac{ 2n_f({\bf p})-1}{E_p}.
\ee
The result holds in the Hard Loop Approximation. For the pseudoscalar $Y_p$ we obtain the same expression because
$\beta^p_{ij}\beta^q_{ji}=-4\delta^{pq}$, $\gamma_5 \gamma_\mu = -\gamma_\mu \gamma_5$ and $\gamma_5^2 = 1$.
Therefore, we replace the indices $p,q$ by $A,B$ and we write down the result (\ref{P-k-a-2}) as 
\be
\label{P-k-a-5}
_{(a)}P^{AB}_{ab}(k) = -4 g^2 N_c \delta_{ab} \delta^{AB}
\int \frac{d^3p}{(2\pi )^3} \, \frac{ 2n_f({\bf p})-1}{E_p}.
\ee

The contribution represented by the graph depicted in Fig.~\ref{fig-scalar}b equals
\ba
\label{P-k-b-1}
{_{(b)}}P^{AB}_{ab}(k) = - i \frac{1}{2} g^2 N_c \delta_{ab} \delta^{AB}
\int \frac{d^4p}{(2\pi)^4}
\big[(p+2k)^2 \Delta^+(p+k) \,  D^{\rm sym}(p)
+ (p+k)^2 \Delta^{\rm sym}(p) \, D^-(p-k) \big] ,
\ea
which after the substitution of the functions $D^{\pm}, \, D^{\rm sym}$
and $\Delta^{\pm}, \Delta^{\rm sym}$ in the form (\ref{D-pm}, \ref{D-sym}, \ref{Del-pm},
\ref{Del-sym}) leads to
\ba
\label{P-k-b-3}
_{(b)}P^{AB}_{ab}(k) & = & \frac{1}{2} \, g^2 N_c \delta_{ab} \delta^{AB}
\int \frac{d^3p}{(2\pi )^3} \frac{4n_g({\bf p}) - 2n_s({\bf p}) + 1}{E_p} 
\ea
within the Hard Loop Approximation.

The contributions coming from the gluon tadpole shown in Fig.~\ref{fig-scalar}c and 
the scalar tadpole from Fig.~\ref{fig-scalar}d equal, respectively, 
\ba
\label{P-k-c-2}
_{(c)}P^{AB}_{ab}(k) &=& - 2 g^2 N_c \delta_{ab} \delta^{AB}
\int \frac{d^3p}{(2\pi)^3} \frac{2n_g({\bf p})+1}{E_p},
\\[2mm]
\label{P-k-d-2}
_{(d)}P^{AB}_{ab}(k) &=& - 5g^2 N_c \delta_{ab} \delta^{AB}
\int \frac{d^3p}{(2\pi)^3} \frac{2n_s({\bf p})+1}{2E_p}.
\ea
In both cases the symmetry factor $1/2$ is included. 

Summing up all contributions we obtain the final formula of scalar self-energy
\be
\label{P-k-final}
P^{AB}_{ab}(k) = - g^2 N_c \delta_{ab} \delta^{AB}
\int \frac{d^3p}{(2\pi)^3} \frac{f({\bf p})}{E_p},
\ee
which depends, as $\Pi$ and $\Sigma$, only on the effective distribution function (\ref{f-def}).

\section{Effective Action}
\label{sec-eff-action}

The Hard Loop Approach can be formulated in an elegant and compact way by introducing the effective action which was first derived for equilibrium plasmas in \cite{Taylor:1990ia,Frenkel:1991ts,Braaten:1991gm} within the thermal field theory. It was also rederived in terms of quasiclassical kinetic theory \cite{Blaizot:1993be,Kelly:1994dh}. Later on a generalization of the action to anisotropic systems was given in \cite{Pisarski:1997cp,Mrowczynski:2004kv}.

A structure of the effective action is constraint by the form of respective self-energies. Since the self-energy of a given field is the second functional derivative of the action with respect to the field, one writes
\ba
\label{action-A-1}
{\cal L}^{A}_2(x) &=&
\frac{1}{2} \int d^4y \; A_\mu^a(x) \Pi_{ab}^{\mu \nu}(x-y) A_\nu^b(y) ,
\\ [2mm]
\label{action-Psi-1}
{\cal L}^{\Psi}_2(x) &=&
\int d^4y \; \bar{\Psi}_i^a(x) \Sigma^{ij}_{ab} (x-y) \Psi_j^b(y) ,
\\ [2mm]
\label{action-Phi-1}
{\cal L}^{\Phi}_2(x) &=&
\frac{1}{2} \int d^4y \; \Phi_A^a(x) P^{AB}_{ab}(x-y) \Phi_B^b(y) ,
\ea
where the self-energies are given by the formulas (\ref{Pi-k-final}, \ref{Si-k-final}, \ref{P-k-final}), respectively. The subscript `2' indicates that the above effective actions generate only two-point functions. To generate $n$-point functions these actions need to be modified to a gauge invariant form. In the nonAbelian gauge theory studied here, the actions (\ref{action-A-1}, \ref{action-Psi-1}, \ref{action-Phi-1}) require a simple change - the ordinary derivatives should be replaced by the covariant ones in the final expressions. Repeating the calculations described in detail in \cite{Mrowczynski:2004kv}, one finds the Hard Loop effective action of the ${\cal N}=4$ super Yang-Mills as
\ba
{\cal L}_{\rm HL}
&=&
-\frac{1}{4}F^{\mu \nu}_a F_{\mu \nu}^a
+\frac{i}{2}\bar \Psi_i^a (D\!\sla \Psi_i)^a
+\frac{1}{2}(D_\mu \Phi_A)_a (D^\mu \Phi_A)_a
\\ \nn
&& + \; {\cal L}^{A}_{\rm HL} +{\cal L}^{\Psi}_{\rm HL} +{\cal L}^{\Phi}_{\rm HL} ,
\ea
where
\ba
\label{action-A-2}
{\cal L}^{A}_{\rm HL} &=& g^2 N_c  \int \frac{d^3p}{(2\pi )^3} \,
\frac{f({\bf p})}{E_p} \,
F_{\mu \nu}^a (x) \bigg({p^\nu p^\rho \over (p \cdot D)^2}\bigg)_{ab} F_\rho^{b \;\mu} (x) ,
\\ [2mm]
\label{action-Psi-2}
{\cal L}^{\Psi}_{\rm HL} &=& g^2 N_c
\int \frac{d^3p}{(2\pi )^3} \, \frac{ f({\bf p})}{E_p} \,
\bar{\Psi}^a_i(x) \bigg( {p \cdot \gamma \over p\cdot D}\bigg)_{ab} \Psi^b_i(x) ,
\\ [2mm]
\label{action-Phi-2}
{\cal L}^{\Phi}_{\rm HL} &=& - \frac{g^2 N_c}{2}
\int \frac{d^3p}{(2\pi )^3} \, \frac{f({\bf p})}{E_p} \;
\Phi_A^a(x) \Phi_A^a(x) .
\ea
where $f({\bf p})$ is, as previously, the  effective distribution function of plasma constituents.

The actions  (\ref{action-A-2}, \ref{action-Psi-2}, \ref{action-Phi-2}) are obtained from the self-energies but the reasoning can be turned around. As argued in \cite{Frenkel:1991ts,Braaten:1991gm}, the actions of gauge bosons (\ref{action-A-2}), fermions (\ref{action-Psi-2}), and scalars (\ref{action-Phi-2}) are of unique gauge invariant form. Therefore, the structures of hard-loop self-energies of gauge bosons, fermions and scalars are unique. Consequently, the self-energies computed in the previous section and those corresponding to the fundamental ${\cal N} = 2$ hypermultiplet can be inferred from the known QED and QCD results with some help of supersymmetry arguments. However,  explicit computations, as those  presented in Sec.~\ref{sec-self-energies}, seem to be still needed to determine, at least, numerical coefficients.

\section{Collective modes}
\label{sec-modes}

When the self-energies computed in Sec.~\ref{sec-self-energies} are substituted into the dispersion equations presented in Sec.~\ref{sec-dis-eqs}, collective modes can be found as solutions of the equations. Below we briefly discuss the gluon, fermion, and scalar excitations.

\begin{itemize}

\item
The structure of polarization tensor (\ref{Pi-k-final}) is such as of a gluon polarization tensor in QCD plasma. It has also an  analogical form as in both usual and supersymmetric QED plasma. Therefore, the spectrum of collective excitations of gauge bosons is in all cases the same. In equilibrium plasma we have the longitudinal (plasmon) mode and the transverse one which are discussed in {\it e.g.} the textbook \cite{lebellac}. When the plasma is out of equilibrium there is a whole variety of possible collective excitations. In particular, there are unstable modes, see  {\it e.g.} the review  \cite{Mrowczynski:2007hb}, which exponentially grow in time and strongly influence the system's dynamics.

\item
The form of Majorana fermion self-energy (\ref{Si-k-final}) happens to be the same as the quark self-energy in QCD plasma. It also coincides with the electron self-energy in both non-supersymmetric and supersymmetric QED plasma. Therefore, we have an identical spectrum of excitations of fermions in all these systems. In equilibrium plasma there are two modes of opposite helicity over chirality ratio, see in {\it e.g.} the textbook \cite{lebellac}.  One mode corresponds to the positive energy fermion, another one, sometimes called a plasmino, is a specific medium effect.  In non-equilibrium plasma the spectrum of fermion collective excitations changes but no unstable modes have been found even for an extremely anisotropic momentum distribution  \cite{Mrowczynski:2001az,Schenke:2006fz}.

\item
The scalar self-energy (\ref{P-k-final}) is  independent of momentum, it is negative and real. Therefore,  $P(k)$ can be written as $P(k) = - m^2_{\rm eff}$ where $m_{\rm eff}$ is the effective scalar mass. Then, the solutions of dispersion equation (\ref{dis-eq-selectron}) are $E_p = \pm \sqrt{m^2_{\rm eff} + {\bf p}^2}$.

\end{itemize}

We conclude this section by saying that the gauge boson and fermion excitations of SYMP are the same as in ultrarelativistic QED and QCD plasma. The scalar excitations are of the form of a free massive relativistic particle.

\section{Collisional characteristics}
\label{sec-collisions}

We consider here characteristics of  the ${\cal N} = 4$ super Yang-Mills plasma which are driven by collisions of plasma constituents.  We start with a review of elementary processes and then we discuss transport coefficients. 

\subsection{Elementary processes}
 
The elementary processes, which occur at the lowest nontrivial order of the coupling constant $g$, are binary interactions, the cross sections of which are proportional to $g^4$. Table~\ref{table-collisions} gives the respective matrix elements squared summed over all internal degrees of freedom of interacting particles. $G, F, S$ denote a gluon, fermion and scalar, respectively. The matrix elements, which were first computed in \cite{Huot:2006ys}, are expressed through the Mandelstam invariants $s,t$ and $u$ defined in the standard way. For a process symbolically denoted as $1+2 \longrightarrow 3+4$, we have
\be
s \equiv (p_1 +p_2)^2 , \;\;\;\;\;\;  t \equiv (p_1 - p_3)^2 , \;\;\;\;\;\;  u \equiv (p_1 - p_4)^2,
\ee
where $p_1, p_2, p_3, p_3$ are the four-momenta of particles $1, 2, 3, 4$, respectively. For a given process, the differential cross section,  which is summed over the internal degrees of freedom of final state particles and averaged over  the  internal degrees of freedom of initial state particles, is expressed through the matrix element squared  from  Table~\ref{table-collisions} as
\be 
\frac{d \sigma}{d t} = \frac{1}{16 \pi s^2}  \frac{1}{N^{\rm dof}_1} \frac{1}{N^{\rm dof}_2}\sum |M|^2,
\ee
where $N^{\rm dof}_1$ and $N^{\rm dof}_2$ are the numbers of internal degrees of freedom of initial state particles  given in Table~\ref{table-field-content}.  The collisional processes listed in Table~\ref{table-collisions} determine transport properties of the plasma.

\begin{table}[b]
\caption{\label{table-collisions} Elementary processes in ${\cal N} =4$ super Yang-Mills plasma.}
\begin{ruledtabular}
\begin{tabular}{ccc}
$n^0$ &  Process & $\frac{1}{g^4} \frac{1}{N_c^2(N_c^2 -1)}\sum |M|^2$
\\[4mm]
1& $GG \leftrightarrow GG$ & $8 \big(\frac{s^2 + u^2}{t^2} + \frac{u^2 + t^2}{s^2}+\frac{t^2 + s^2}{u^2} +3 \big)$
\\[2mm]
2 & $GF \leftrightarrow GF$ &  $32 \big(\frac{s^2 + u^2}{t^2} - \frac{u}{s} - \frac{s}{u} \big)$
\\[2mm]
3 & $GG \leftrightarrow FF$ &  $32 \big(\frac{t^2 + u^2}{s^2} - \frac{u}{t} - \frac{t}{u} \big)$
\\[2mm]
4 & $GS \leftrightarrow GS$ & $24 \big(\frac{s^2 + u^2}{t^2} + 1 \big)$
\\[2mm]
5 & $GG \leftrightarrow SS$ &  $24 \big(\frac{t^2 + u^2}{s^2} + 1 \big)$
\\[2mm]
6 & $GF \leftrightarrow SF$ &  $- 96 \big( \frac{u}{s} + \frac{s}{u} +1 \big)$
\\[2mm]
7 & $GS \leftrightarrow FF$ & $- 96 \big( \frac{u}{t} + \frac{t}{u} +1 \big)$
\\[2mm]
8 & $FS \leftrightarrow FS$ &  $- 96 \big[ \frac{2us}{t^2} + 3 \big( \frac{u}{s} + \frac{s}{u} \big) + 1 \big]$
\\[2mm]
9 & $SS \leftrightarrow FF$ & $- 96 \big[ \frac{2ut}{s^2} + 3 \big( \frac{u}{t} + \frac{t}{u} \big) + 1 \big]$
\\[2mm]
10 & $SS \leftrightarrow SS$ & $72 \big(\frac{s^2 + u^2}{t^2} + \frac{u^2 + t^2}{s^2}+\frac{t^2 + s^2}{u^2} +3 \big)$
\\[2mm]
11 & $FF \leftrightarrow FF$ & $128 \big(\frac{s^2 + u^2}{t^2} + \frac{u^2 + t^2}{s^2}+\frac{t^2 + s^2}{u^2} +3 \big)$
\\[2mm]
\end{tabular}
\end{ruledtabular}
\end{table}

\subsection{Transport coefficients}

Transport coefficients of weakly coupled QGP, which include baryon and strangeness diffusion, electric charge and heat conductivity, shear and bulk viscosity and color conductivity, have been studied in detail, see \cite{Arnold:2000dr,Arnold:2003zc,Arnold:2006fz,Arnold:1998cy} and references therein. The shear viscosity of SYMP has been computed in \cite{Huot:2006ys} and the bulk viscosity  is identically zero because of exact conformality of the system. Other transport coefficients of SYMP have not been studied but one expects the coefficients to be qualitatively similar to those of QGP.

Since the temperature is the only dimensional parameter, which characterizes the equilibrium plasma of massless constituents, one finds that, for example, the shear viscosity $\eta$ must be proportional to $T^3$ and the color conductivity  $\sigma_c$ to $T$. It appears that the dominant contributions to both transport coefficients of QGP come from the binary collisions driven by a one-gluon exchange which correspond to the matrix elements squared diverging as $t^{-2}$ for $t \rightarrow 0$. The analyses presented in \cite{Arnold:2000dr} and \cite{Arnold:1998cy}, respectively, show that at the leading order $\eta \sim T^3/g^4\ln g^{-1}$ and $\sigma_c \sim T/\ln g^{-1}$. The factor $1/\ln g^{-1}$ appears due to the infrared singularity of the Coulomb-like interaction which is regulated by the gluon self-energy. Actually the physics behind the two formulas is rather different. The viscosity is governed by collisions with the momentum transfer of the order of $gT$ while for the color conductivity the softer collisions with the momentum  transfer of the order $g^2T$ play a crucial role. 

One expects the same parametric form of $\eta$, $\sigma$ and other transport coefficients in the case of SYMP and QGP because, similarly to QGP, there are the Coulomb-like binary interactions for every constituent of SYMP, see Table~\ref{table-collisions}. The analysis \cite{Huot:2006ys} indeed proves that the shear viscosity coefficients of QGP and SYMP differ only by numerical factors which mostly reflect different numbers of degrees of freedom in the two plasmas. The viscosity is strongly dominated by the Coulomb-like interactions, and consequently it does not much matter that the sets of elementary processes in the two plasma systems are different. 

In the paper \cite{Czajka:2011zn} we considered two transport characteristics of the ${\cal N} =1$ QED plasma which are not so constrained by dimensional arguments and seemed to strongly depend on the elementary process under consideration. Specifically, we computed the collisional energy loss and momentum broadening of a particle traversing the equilibrium plasma.  The latter quantity determines a magnitude of radiative energy loss of a highly energetic particle in a plasma \cite{Baier:1996sk}. The dimensional argument does not work here because the two quantities depend not only on the plasma temperature but on the energy of the test particle as well. We computed the energy loss and momentum broadening due to the processes which, like the Compton scattering on selectrons, are independent of momentum transfer. Such processes are qualitatively different from the Coulomb-like interactions dominated by small momentum transfers.  We managed to obtain the exact formulas of the energy loss and momentum broadening due to the momentum-independent scattering. In the limit of the high energy of test particle, which is important in the context of jet suppression phenomenology in nucleus-nucleus collisions, the energy loss and momentum broadening appeared to be very similar (at the leading order) to those driven by the Coulomb-like interactions. 

The result can be understood as follows. One estimates the energy loss  $\frac{dE}{d x}$ as $\langle \Delta E \rangle / \lambda$, where $\langle \Delta E \rangle$ is the typical change of a particle's energy in a single collision and $\lambda$ is the particle's mean free path given as $\lambda^{-1} = \rho \, \sigma$ with $\rho \sim T^3$ being the density of scatterers and $\sigma$ denoting the cross section. For the differential cross section, which is independent of momentum transfer, the total cross section is $\sigma \sim e^4/s$. When a highly energetic particle with energy $E$ scatters on massless plasma particle, $s \sim ET$ and consequently  $\sigma \sim e^4/(ET)$. The inverse mean free path is thus estimated as $\lambda^{-1} \sim e^4 T^2/E$.  When the scattering process is independent of momentum transfer, $\langle \Delta E \rangle$ is of order $E$ and we finally find $-\frac{dE}{d x} \sim e^4 T^2$. In the case of Coulomb interaction we have $\langle \Delta E \rangle \sim - e^2 T$, $\lambda^{-1} = e^2 T$ which provide the same estimate of the energy loss. The energy transfer in a single collision is thus much smaller in the Coulomb interaction than in the momentum independent scattering but the cross section is bigger in the same proportion.  Consequently, the two interactions corresponding to very different differential cross sections lead to very similar energy losses. 

We expect an analogous situation in SYMP. There are various elementary process but the energy loss and momentum broadening of highly energetic particles do not much differ from those in QGP.

\section{Conclusions}
\label{sec-conclusions}

QCD is obviously rather different from ${\cal N} = 4$ super Yang-Mills theory. Nevertheless QGP and SYMP are surprisingly similar in the weak coupling regime (at the leading order). The form of gluon collective excitations is identical and the same is true for the fermion (quark) modes.  The scalar modes in SYMP are as of a massive relativistic particle. The sets of elementary processes are different in QGP and SYMP but the transport coefficients, which are dominated by the Coulomb-like interactions, are quite similar.  The energy loss and momentum broadening of a highly energetic test particle are also rather similar in the two plasma systems. The differences mostly come from different numbers of degrees of freedom in both plasmas which need to be taken into account for a quantitative comparison.

\section*{Acknowledgments}
We are grateful to Simon Caron-Huot for a discussion on conserved charges in SYMP. This work was partially supported by the ESF Human Capital Operational Program and Polish Ministry of Science and Higher  Education under grants  6/1/8.2.1/POKL/2009 and 667/N-CERN/2010/0, respectively.

\newpage
\appendix

\section{Vertexes of ${\cal N}=4$ Super Yang-Mills}
\label{sec-SYM-vertex}

We collect here the vertex functions which are inferred from the Lagrangian (\ref{Lagrangian-1}). Since all fields of the ${\cal N}=4$ super Yang-Mills are, except the ghosts, real, there are no arrows orienting the lines. However, one should remember that the momentum of every gluon in the three-gluon coupling is assumed to enter the vertex.  In the case of the gluon coupling to scalars, the momentum of one scalar enters the vertex and the momentum of the other one leaves it.

\begin{table}[!h]
\begin{ruledtabular}
\begin{tabular}{m{2cm} m{.2cm} m{5cm} m{.2cm} m{9cm}}
        \\
        \centering 1 &&
        \centering \includegraphics[scale=.08]{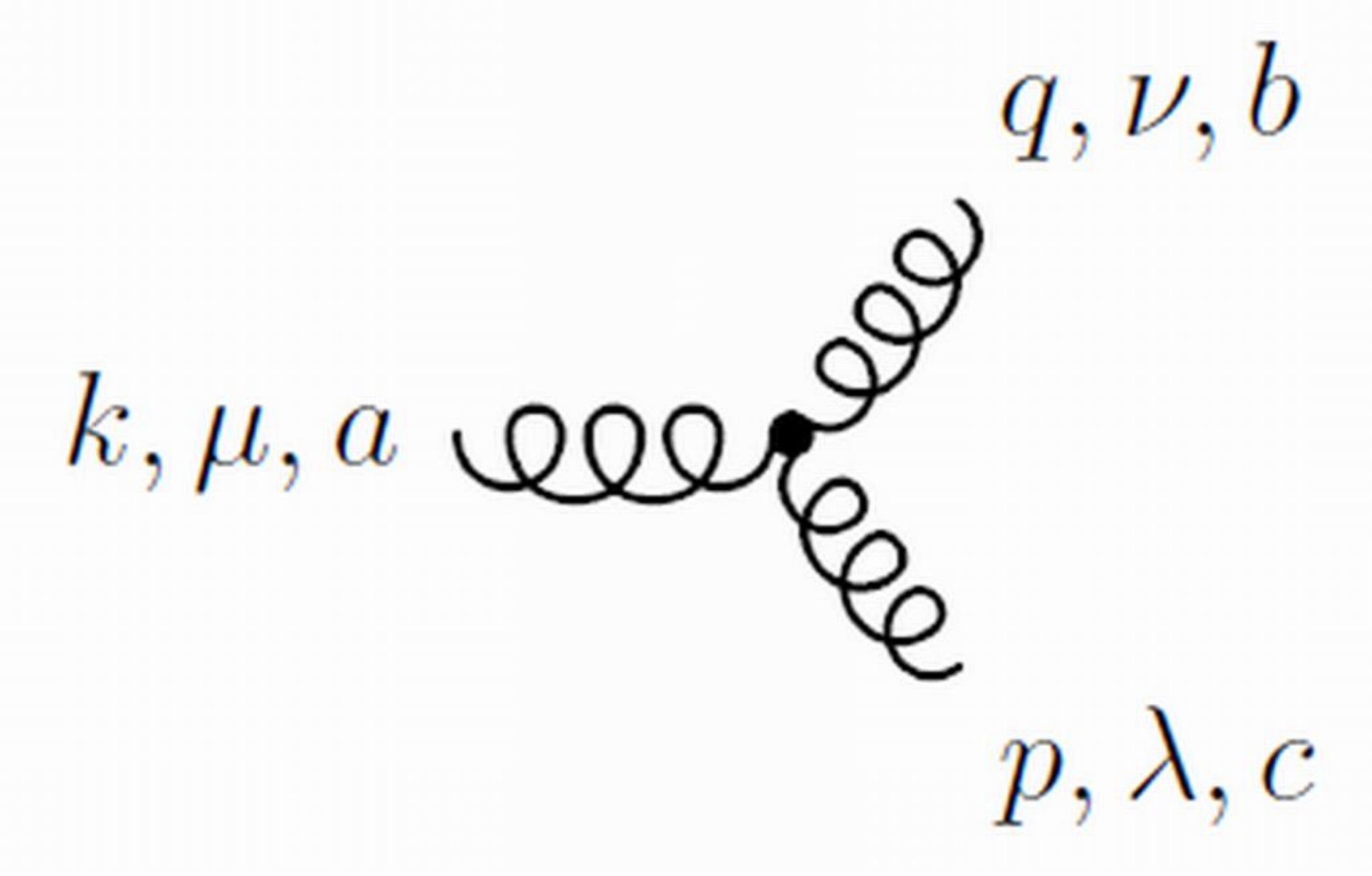}&&
        \ba -gf^{abc} \big[g^{\mu\nu}(k-q)^\lambda + g^{\nu\lambda}(q-p)^\mu+g^{\lambda\mu}(p-k)^\nu \big] \nn \ea
        \\
        \centering 2 &&
        \centering \includegraphics[scale=.06]{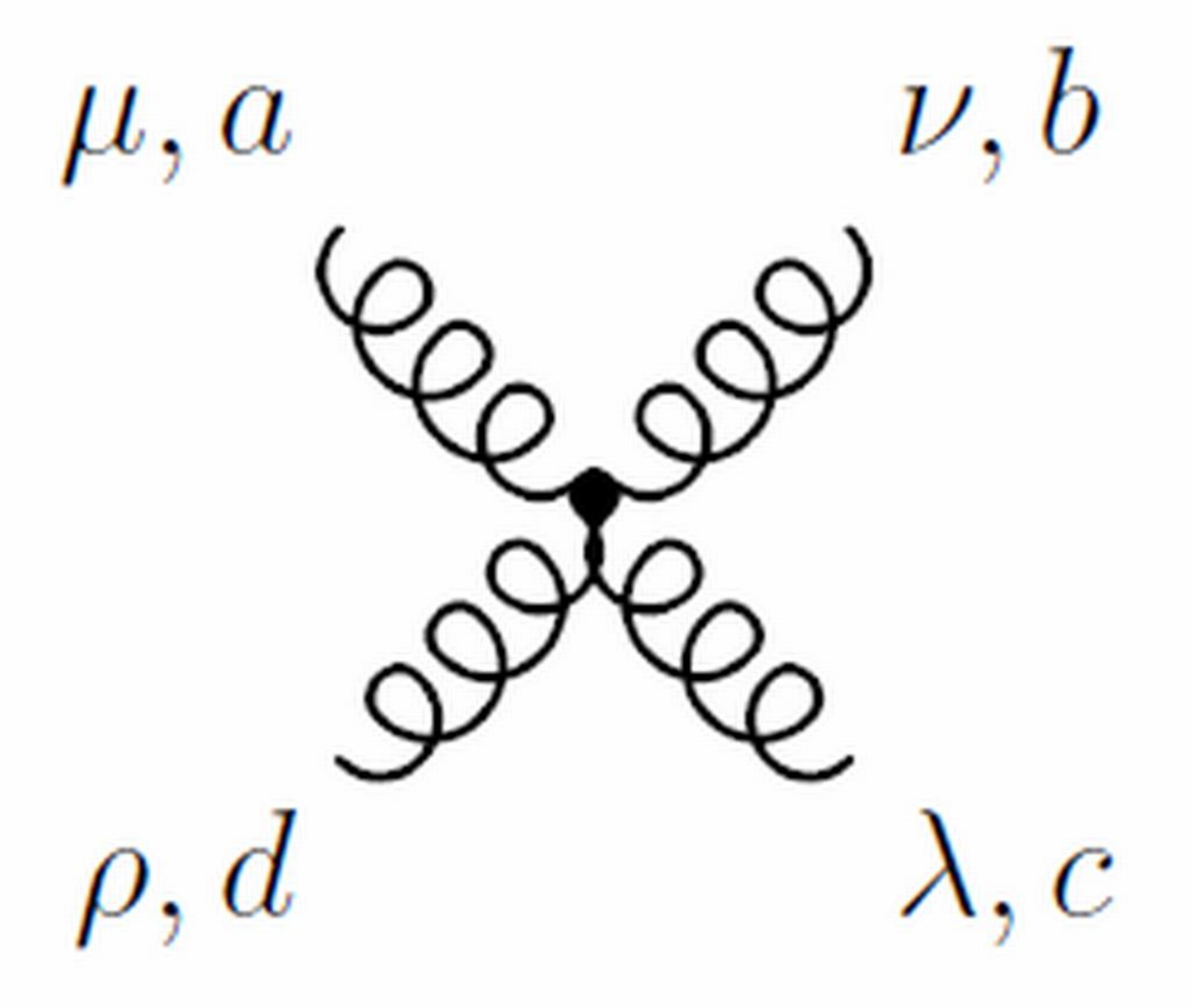} &&
         \ba &-ig^2& \big[\; f^{abe}f^{cde}(g^{\mu\lambda}g^{\nu\rho}-g^{\mu\rho}g^{\nu\lambda}) \nn \\
         && +f^{ace}f^{bde}(g^{\mu\nu}g^{\lambda\rho}-g^{\mu\rho}g^{\nu\lambda}) \nn \\
         && +f^{ade}f^{cbe}(g^{\mu\lambda}g^{\nu\rho}-g^{\mu\nu}g^{\rho\lambda})\;\big] \nn \ea
         \\
        \centering 3 &&
        \centering \includegraphics[scale=.08]{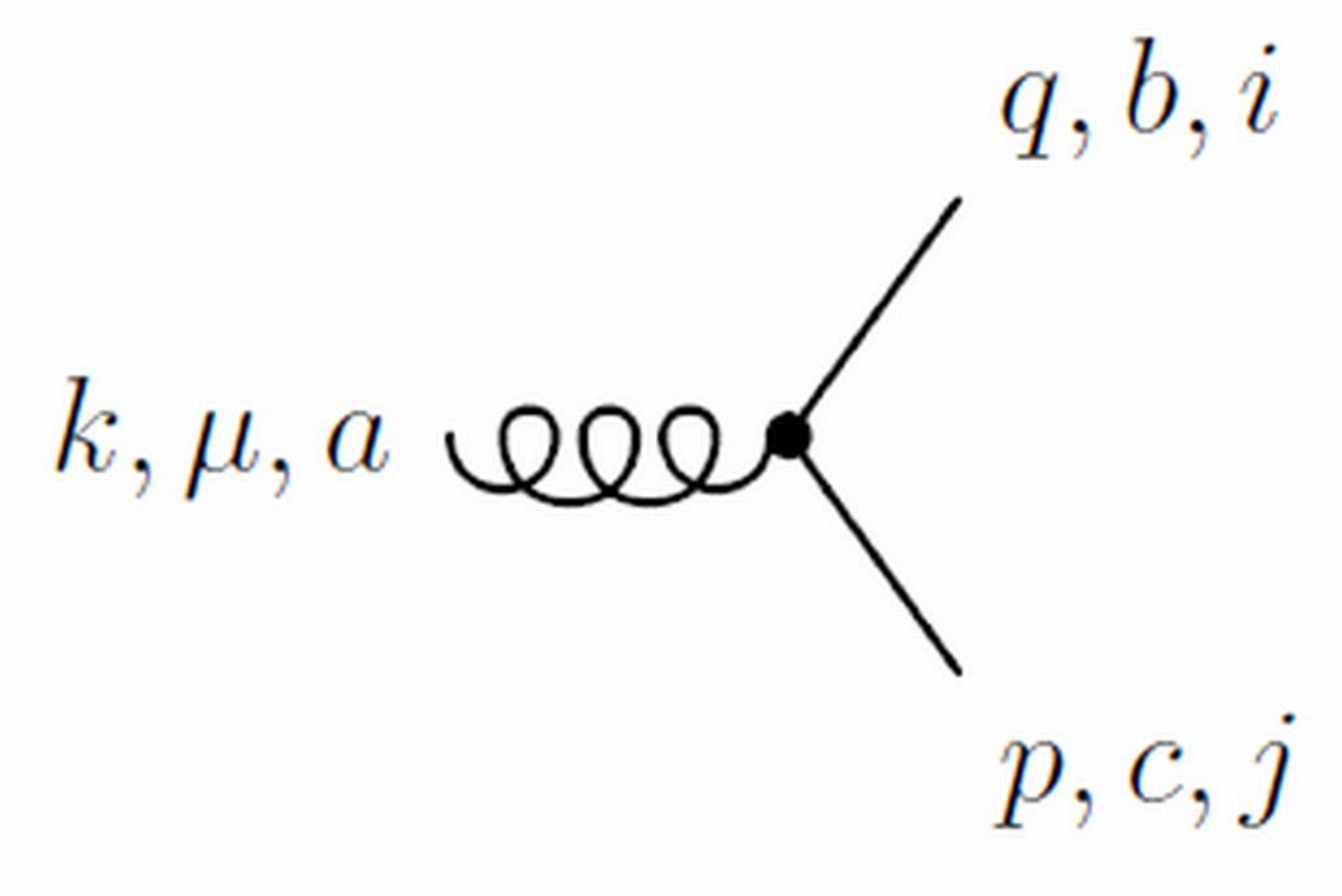} &&
        \ba gf^{abc}\delta^{ij}\gamma^\mu \nn \ea
        \\
        \centering  4 &&
        \centering \includegraphics[scale=.08]{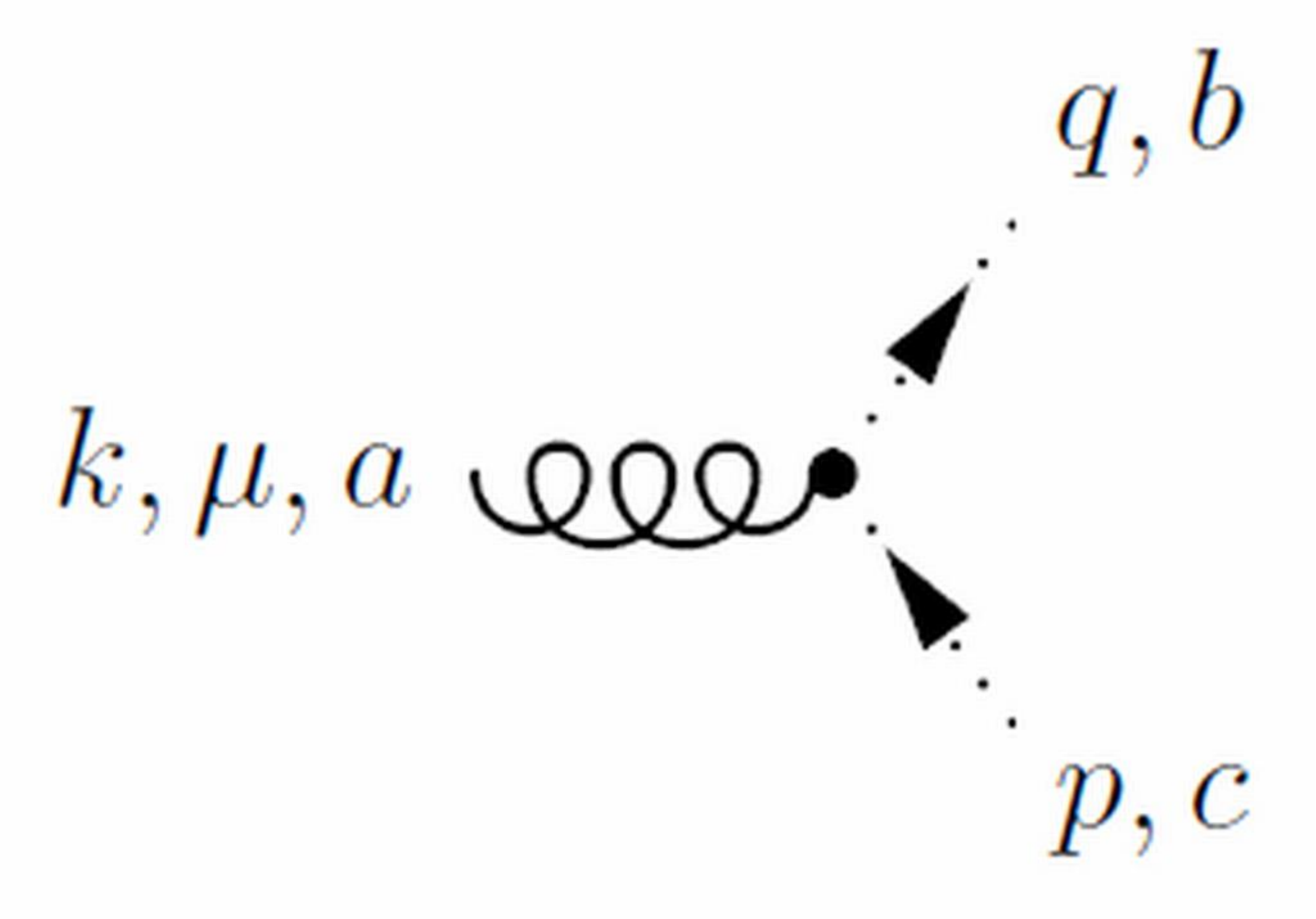} &&
        \ba gf^{abc}q_\mu \nn \ea
         \\ 
        \centering  5 &&
        \centering \includegraphics[scale=.08]{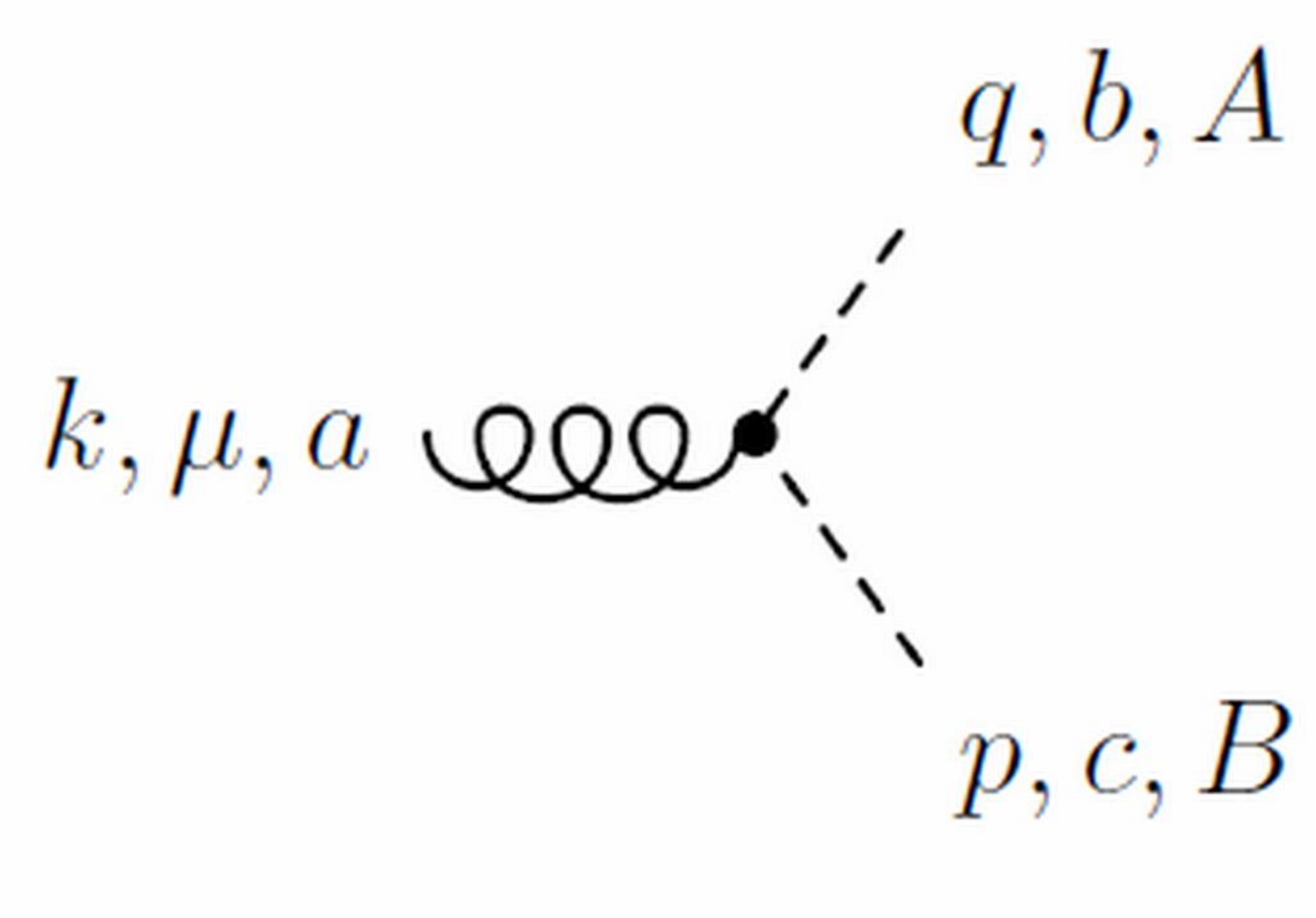} &&
        \ba gf^{abc}\delta^{AB} (p+q)_\mu \nn \ea
         \\ 
       \centering  6 &&
       \centering \includegraphics[scale=.08]{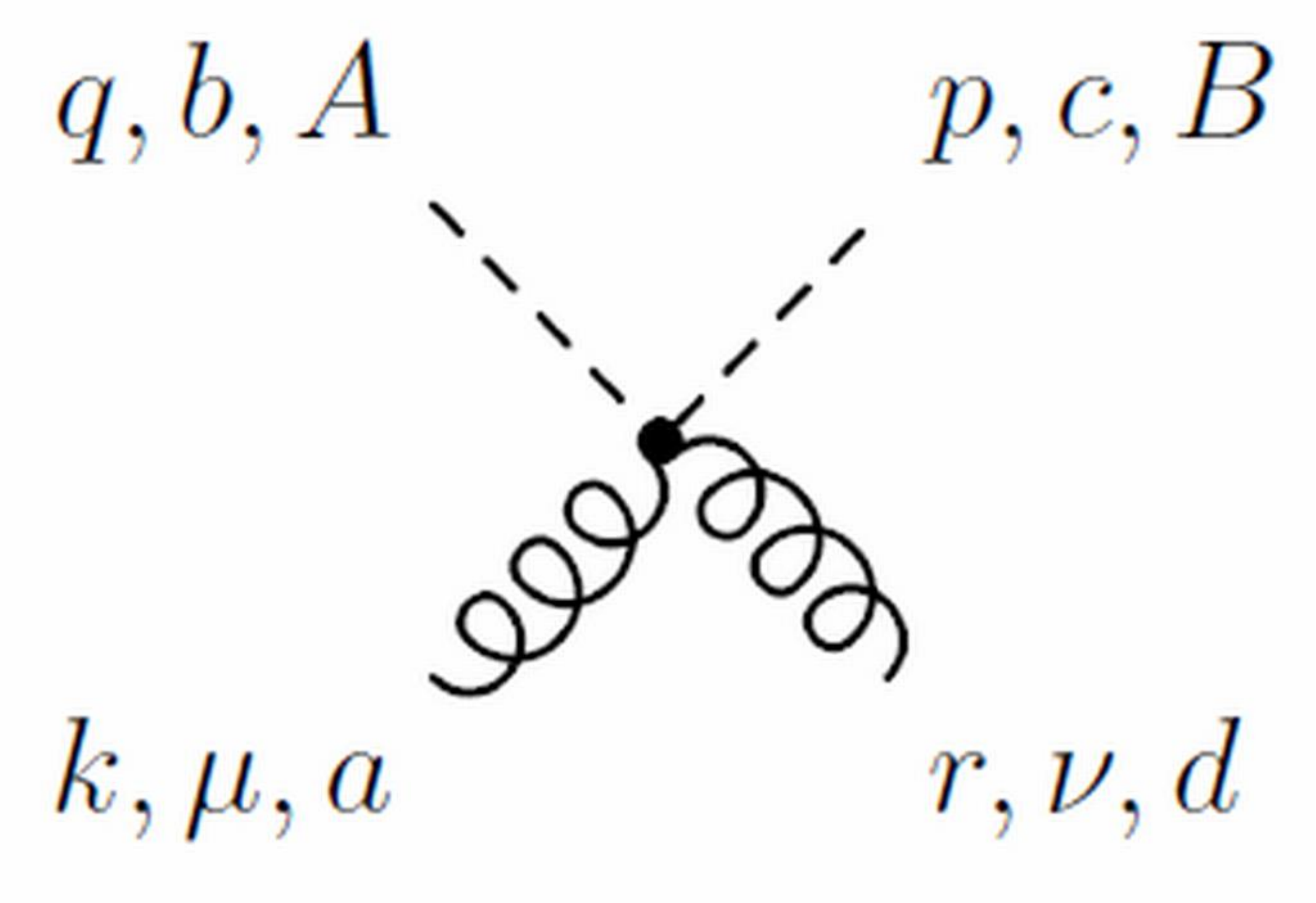} &&
        \ba 2ig^2 g^{\mu\nu} f^{abe}f^{cde}\delta^{AB} \nn \ea
         \\
        \centering  7 &&
        \centering \includegraphics[scale=.08]{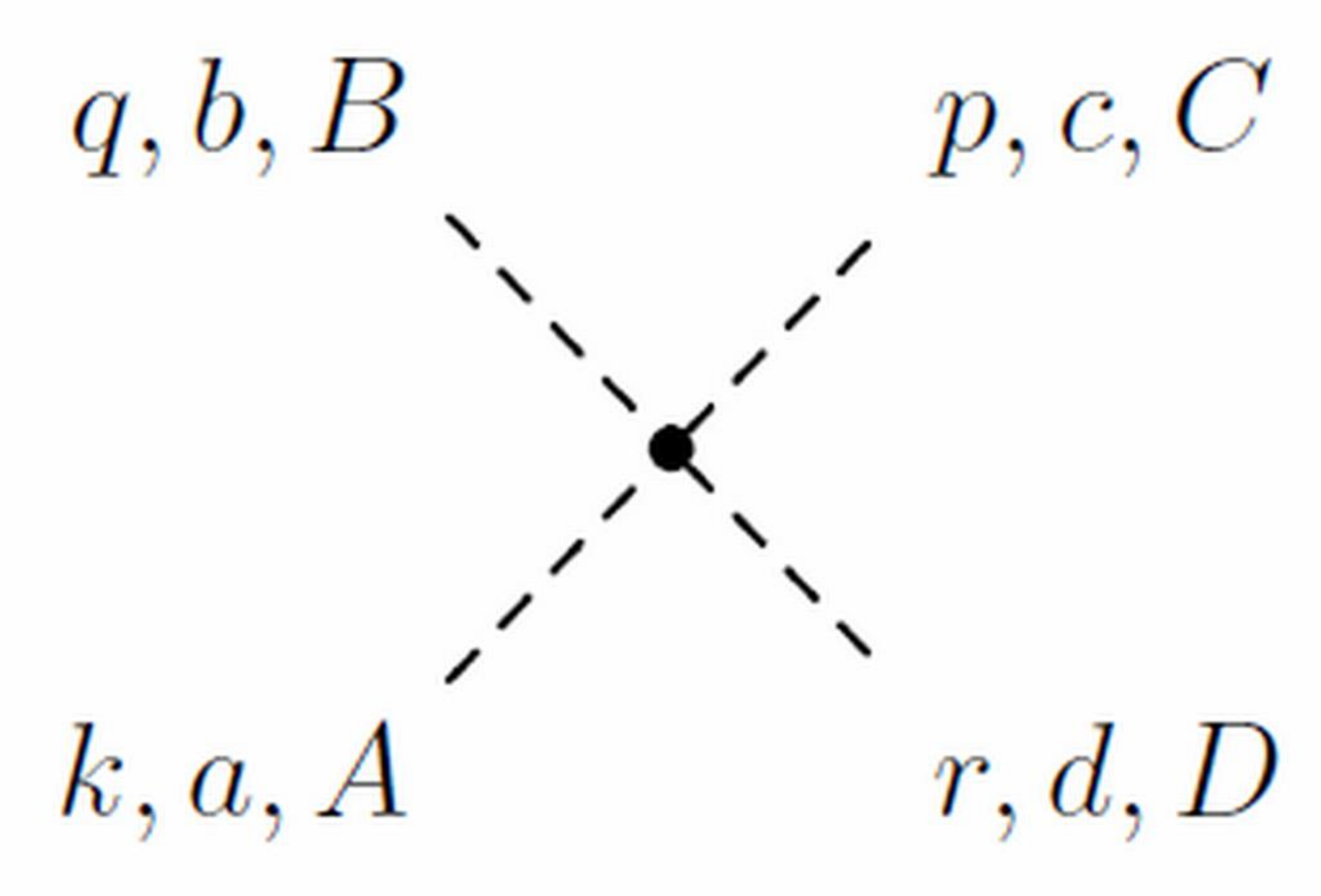}&&\ba
        &-ig^2&\big[\;f^{abe}f^{cde}(\delta^{AC}\delta^{BD}-\delta^{AD}\delta^{BC})\nn \\
        &&+f^{ace}f^{bde}(\delta^{AB}\delta^{CD}-\delta^{AD}\delta^{BC})\nn \\
        &&+f^{ade}f^{cbe}(\delta^{AD}\delta^{CB}-\delta^{AB}\delta^{DC})\;\big]\nn \ea
         \\
        \centering  8 &&
        \centering \includegraphics[scale=.08]{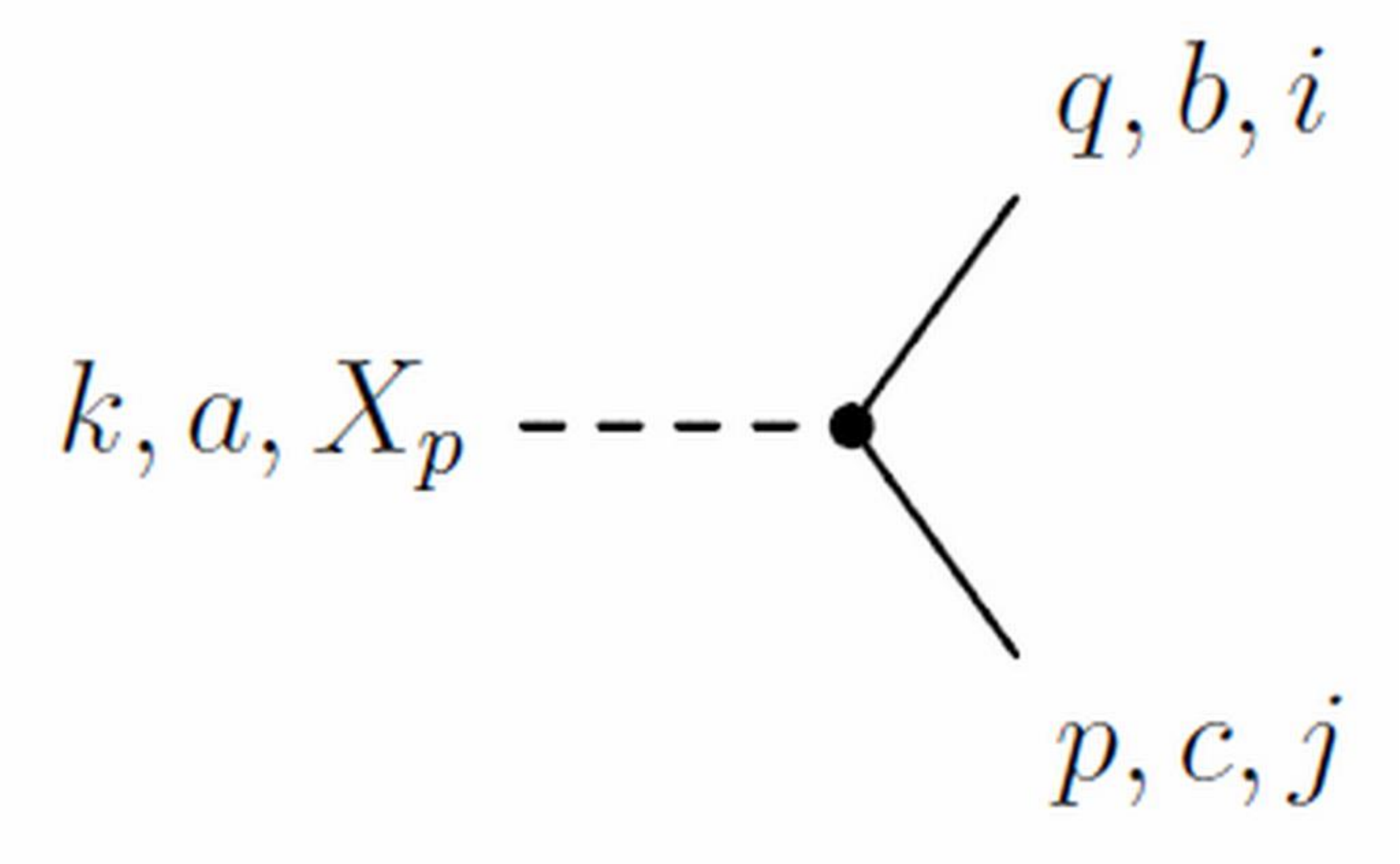} &&
        \ba -ig f^{abc} \alpha^p_{ij} \nn \ea
         \\ 
        \centering  9 &&
        \centering \includegraphics[scale=.08]{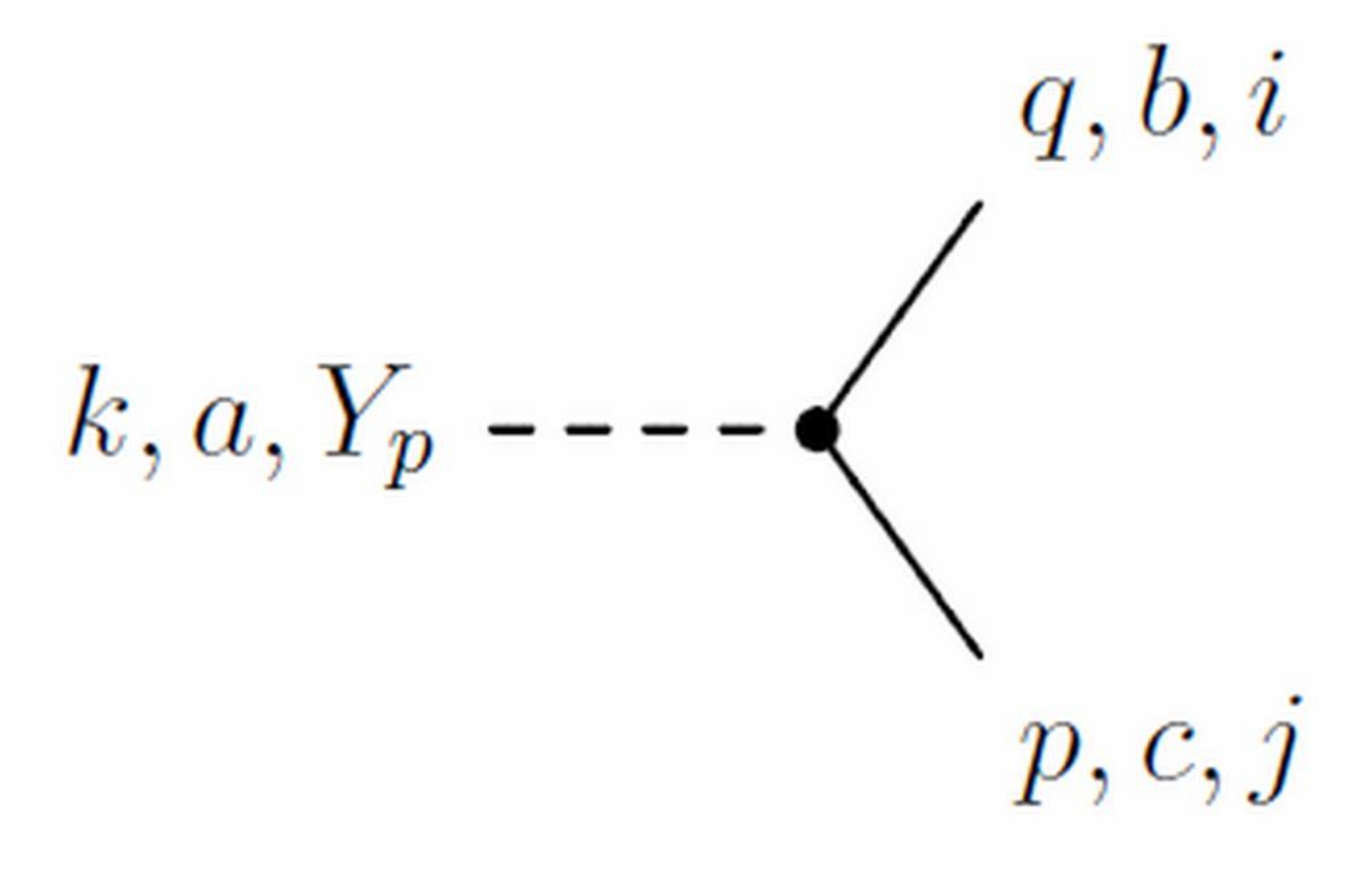} &&
        \ba gf^{abc} \beta^p_{ij}\gamma_5 \nn \ea
         \\
\end{tabular}
\end{ruledtabular}
\end{table}

\section{Green's functions of $N=4$ super Yang-Mills}
\label{sec-Green-fun}

We present here the retarded, advanced and unordered free Green's functions of the Keldysh-Schwinger formalism which are usually labeled with the indices $+, -, >, <$, respectively. The system is assumed to be translationally invariant and locally colorless. The functions of gluons are given in the Feynman gauge.

\subsection{Gluons}

The functions of interest of the free gluon field are of the form
\ba
\label{D-pm}
(D^{\pm}(p))^{\mu \nu}_{ab} &=& g^{\mu \nu} \delta_{ab} D^{\pm}(p)
= -\frac{g^{\mu \nu} \delta_{ab}}{p^2\pm i\, {\rm sgn}(p_0)0^+},
\\
\label{D->}
(D^>(p))^{\mu \nu}_{ab} &=& g^{\mu \nu} \delta_{ab} D^>(p)
= g^{\mu \nu} \delta_{ab}\frac{i\pi}{E_p}
\Big(\delta (E_p - p_0) \big[ n_g ({\bf p}) +1\big]
+ \delta (E_p + p_0) n_g (-{\bf p}) \Big),
\\
\label{D-<}
(D^< (p))^{\mu \nu}_{ab} &=& g^{\mu \nu} \delta_{ab} D^<(p)
= g^{\mu \nu} \delta_{ab}\frac{i\pi }{E_p}
\Big( \delta (E_p - p_0) n_g ({\bf p})
+ \delta (E_p + p_0)  \big[ n_g (-{\bf p}) + 1\big] \Big),
\\
\label{D-sym}
(D^{\rm sym} (p))^{\mu \nu}_{ab} &\equiv& (D^>(p))^{\mu \nu}_{ab} + (D^<(p))^{\mu \nu}_{ab}
= g^{\mu \nu} \delta_{ab} D^{\rm sym}(p) \nn \\
&=& g^{\mu \nu} \delta_{ab}\frac{i\pi }{E_p}
\Big( \delta (E_p - p_0) \big[2 n_g({\bf p}) +1\big]
+ \delta (E_p + p_0) \big[2 n_g(-{\bf p})+ 1\big]\Big),
\ea
where $E_p \equiv |{\bf p}|$ and $n_g({\bf p})$ is the distribution function of gluons which are assumed to be unpolarized. The function is normalized in such a way that the gluon density is given as
\be
\rho_g = 2 (N_c^2 -1)  \int \frac{d^3p}{(2\pi)^3}\, n_g ({\bf p}) ,
\ee
where the factor of 2 takes into account two gluon spin states.

One checks that the functions (\ref{D-pm}, \ref{D->}, \ref{D-<}) obey the required identity
\be
\label{id-D}
D^>(p) - D^<(p) = D^+(p) - D^-(p) .
\ee
Indeed, the left-hand side of Eq.~(\ref{id-D}) equals
\be
\label{D>-D<}
(D^>(p))^{\mu \nu}_{ab} - (D^< (p))^{\mu \nu}_{ab} = \frac{i\pi g^{\mu \nu} \delta_{ab}}{E_p}
\big(\delta (E_p - p_0)  - \delta (E_p + p_0)  \big)
=2i \pi g^{\mu \nu} \delta_{ab} \delta (p^2) \big(\Theta(p_0)  - \Theta(- p_0) \big).
\ee
Using the well-known relation
\be
\frac{1}{x \pm i0^+} = {\cal P}\frac{1}{x } \mp i \pi \delta(x),
\ee
one immediately shows that the right-hand side of Eq.~(\ref{id-D}) equals
the expression (\ref{D>-D<}).

\subsection{Ghosts}

The functions of the free ghost field are 
\ba
\label{G-pm}
G^{\pm}_{ab}(p) &=& \delta_{ab} G^{\pm}(p)
= -\frac{\delta_{ab}}{p^2\pm i\, {\rm sgn}(p_0)0^+},
\\
\label{G->}
G^>_{ab}(p) &=& \delta_{ab}G^{>}(p)
= \delta_{ab} \frac{i\pi}{E_p} \Big( \delta (E_p - p_0) \big[
n_g({\bf p}) +1\big] +  \delta (E_p + p_0) n_g (-{\bf p})
\Big) ,
\\
\label{G-<}
G^<_{ab}(p) &=& \delta_{ab} G^{<}(p)
= \delta_{ab} \frac{i\pi}{E_p} \Big( \delta
(E_p - p_0) n_g({\bf p}) +  \delta (E_p + p_0)  \big[ n_g(-{\bf p}) + 1\big] \Big) ,
\\
\label{G-sym}
G^{\rm sym}_{ab}(p) &\equiv& G^>_{ab}(p) + G^<_{ab}(p) = \delta_{ab} G^{\rm sym}(p) \nn \\
&=& \delta_{ab} \frac{i\pi }{E_p}\Big( \delta (E_p - p_0) \big[2 n_g({\bf p}) +1\big]
+  \delta (E_p + p_0) \big[2 n_g(-{\bf p}) + 1\big] \Big),
\ea
where $n_g({\bf p})$ is the distribution function of gluons. The functions (\ref{G-pm}, \ref{G->}, \ref{G-<}) obey the identity
analogous to the relation (\ref{id-D}). 

\subsection{Fermions}

The Green's functions of the free massless fermion field equal
\ba
\label{S-pm}
S^{\pm}_{ij}(p)&=& \delta_{ij} S^{\pm}(p)
= \frac{\delta_{ij} {p\sla}}{p^2\pm i\, {\rm sgn}(p_0)0^+},
\\
\label{S->}
S^>_{ij}(p)&=& \delta_{ij} S^{>}(p)
= \delta_{ij} \frac{i\pi}{E_p} {p\sla}
\Big( \delta (E_p - p_0)  \big[ n_f ({\bf p}) -1\big]
+ \delta (E_p + p_0) n_f (-{\bf p}) \Big),
\\
\label{S-<}
S^<_{ij}(p)&=& \delta_{ij} S^{<}(p)
= \delta_{ij} \frac{i\pi}{E_p} {p\sla} \Big( \delta (E_p - p_0)  n_f({\bf p})
+ \delta (E_p + p_0) \big[n_f (-{\bf p}) - 1\big] \Big),
\\
\label{S-sym}
S^{\rm sym}_{ij}(p)&\equiv& S^>_{ij}(p) + S^<_{ij}(p) = \delta_{ij} S^{\rm sym}(p)\nn \\
&=& \delta_{ij} \frac{i\pi}{E_p} {p\sla}
\Big( \delta (E_p - p_0)   \big[ 2n_f({\bf p}) -1 \big]
+ \delta (E_p + p_0) \big[ 2n_f (-{\bf p}) - 1\big] \Big).
\ea
where $n_f({\bf p})$ is the distribution function of fermions. The distribution function is normalized in such a way that the fermion density equals
\be
\rho_f = 2(N_c^2-1) \int \frac{d^3p}{(2\pi)^3}\, n_f({\bf p}) ,
\ee
where the factor of 2 takes into account two spin states of each fermion. The functions (\ref{S-pm}, \ref{S->}, \ref{S-<}) are checked to obey the identity $S^>(p) - S^< (p) = S^+(p) - S^-(p)$.

\subsection{Scalars}
The Green's functions of the free massless scalar field are 
\ba
\label{Del-pm}
\Delta^{\pm}_{ab}(p) &=& \delta_{ab} \Delta^{\pm}(p)
=\frac{\delta^{ab}}{p^2\pm i\, {\rm sgn}(p_0)0^+},
\\
\label{Del->}
\Delta^>_{ab}(p) &=& \delta_{ab} \Delta^{>}(p)
= -\delta_{ab}\frac{i\pi}{E_p} \Big( \delta (E_p - p_0) \big[
n_s({\bf p}) +1\big] +  \delta (E_p + p_0) n_s(-{\bf p})
\Big) ,
\\
\label{Del-<}
\Delta^<_{ab}(p) &=& \delta_{ab} \Delta^{<}(p)
= -\delta_{ab}\frac{i\pi}{E_p} \Big( \delta
(E_p - p_0) n_s({\bf p}) +  \delta (E_p + p_0)  \big[
n_s(-{\bf p}) + 1\big] \Big) ,
\\
\label{Del-sym}
\Delta^{\rm sym}_{ab}(p) &\equiv& \Delta^>_{ab}(p) + \Delta^<_{ab}(p)
= \delta_{ab} \Delta^{\rm sym}(p) \nn \\
&=& - \delta_{ab}\frac{i\pi}{E_p}\Big( \delta (E_p - p_0) \big[2 n_s({\bf p}) +1\big]
+  \delta (E_p + p_0) \big[2 n_s(-{\bf p})+ 1\big] \Big),
\ea
where $n_s({\bf p})$ is the distribution function of scalars. The function is normalized in such
a way that the scalar density equals
\be
\rho_s = (N_c^2 -1) \int \frac{d^3p}{(2\pi)^3}\, n_s({\bf p}) .
\ee
The functions (\ref{Del-pm}, \ref{Del->}, \ref{Del-<}) obey the identity such as the relation (\ref{id-D}).


\end{document}